\documentclass[sigconf]{acmart}

\usepackage{enumitem}
\usepackage[ruled, vlined, linesnumbered, commentsnumbered, longend]{algorithm2e}
\usepackage[noend]{algpseudocode} 
\usepackage{graphicx}
\usepackage{caption}
\usepackage{subfigure}
\usepackage{textcomp}
\usepackage{xcolor}
\usepackage{comment}
\usepackage{flushend}
\usepackage{array}
\usepackage{balance}

\usepackage{multirow}
\usepackage{threeparttable}
\usepackage{soul}
\usepackage{tabularx}
\usepackage{booktabs}
\usepackage{makecell}
\newcolumntype{Y}{>{\centering\arraybackslash}X}
\usepackage{array}
\usepackage{pifont}
\usepackage{color, colortbl}
\definecolor{LightCyan}{RGB}{230, 230, 255}

\usepackage{xcolor}

\usepackage{hyperref}
\hypersetup{colorlinks,linkcolor=yellow,citecolor=blue}
\usepackage[noabbrev]{cleveref}
\crefname{section}{}{\S\S}
\crefdefaultlabelformat{\S#2#1#3}

\usepackage[shortcuts]{extdash}
\usepackage[font=small]{caption}

\usepackage{xspace}
\newcommand{\sys}{Moby\xspace}
\newcommand{\lidar}{LiDAR\xspace}

\AtBeginDocument{%
  }

\copyrightyear{2023}
\acmYear{2023}
\setcopyright{acmlicensed}\acmConference[MM '23]{Proceedings of the 31st ACM International Conference on Multimedia}{October 29-November 3, 2023}{Ottawa, ON, Canada}
\acmBooktitle{Proceedings of the 31st ACM International Conference on Multimedia (MM '23), October 29-November 3, 2023, Ottawa, ON, Canada}
\acmPrice{15.00}
\acmDOI{10.1145/3581783.3612158}
\acmISBN{979-8-4007-0108-5/23/10}

\begin{document}

\title{\sys: Empowering 2D Models for Efficient Point Cloud Analytics on the Edge}

  
  \author{Jingzong Li}
  \affiliation{%
    \country{City University of Hong Kong}
  }
  
  \author{Yik Hong Cai}
  \affiliation{%
    \country{The Chinese University of Hong Kong}
  }

  \author{Libin Liu}
  \affiliation{%
    \country{Zhongguancun Laboratory}
  }
  
  \author{Yu Mao}
  \affiliation{%
    \country{City University of Hong Kong}
  }
  
  \author{Chun Jason Xue}
  \affiliation{%
    \country{City University of Hong Kong}
  }
  
  \author{Hong Xu}
  \affiliation{%
    \country{The Chinese University of Hong Kong}
  }
  
  \renewcommand{\shortauthors}{Jingzong Li et al.}


  \newcommand\blfootnote[1]{%
    \begingroup
    \renewcommand\thefootnote{}\footnote{#1}%
    \addtocounter{footnote}{-1}%
    \endgroup
  }

\begin{abstract}
  3D object detection plays a pivotal role in many applications, most notably autonomous driving and robotics. These applications are commonly deployed on edge devices to promptly interact with the environment, and often require near real-time response. With limited computation power, it is challenging to execute 3D detection on the edge using highly complex neural networks. 
  Common approaches such as offloading to the cloud induce significant latency overheads due to the large amount of point cloud data during transmission. 
  To resolve the tension between wimpy edge devices and compute-intensive inference workloads, we explore the possibility of empowering fast 2D detection to extrapolate 3D bounding boxes. 
  To this end, we present \sys, a novel system that demonstrates the feasibility and potential of our approach.
  We design a transformation pipeline for \sys that generates 3D bounding boxes efficiently and accurately based on 2D detection results without running 3D detectors.
  Further, we devise a frame offloading scheduler that decides when to launch the 3D detector judiciously in the cloud to avoid the errors from accumulating.
  Extensive evaluations on NVIDIA Jetson TX2 with real-world autonomous driving datasets demonstrate that \sys offers up to 91.9\% latency improvement with modest accuracy loss over state of the art.
  \blfootnote{
    This work is supported in part by funding from the Research Grants Council of Hong Kong (11209520, C7004-22G), CUHK (4937007, 4937008, 5501329, 5501517), and Natural Science Foundation of Shandong Province (ZR2022QF070).
  }
\end{abstract}




\maketitle


\section{Introduction}
\label{sec:introduction}

\begin{figure}[t]
	\centering
    \includegraphics[width=0.45\textwidth]{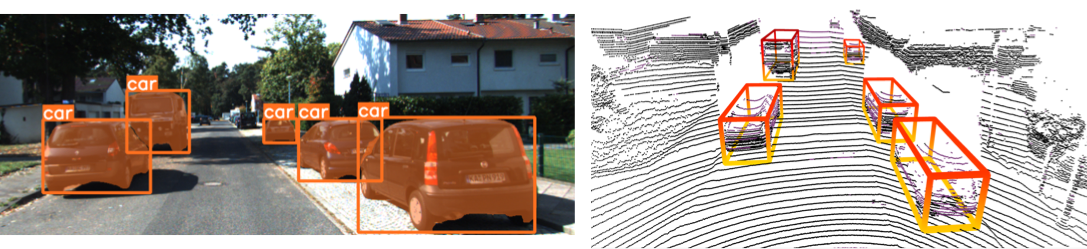}
    \caption{2D and 3D object detection.}
    \label{fig:2d_vs_3d}
    \vspace{-4mm}
\end{figure}

The rapid development of Deep Neural Networks (DNNs) has empowered a number of use cases, including object detection, face recognition and image super-resolution \cite{du2020server, yi2020eagleeye, dong2015image}.
One of the most promising use cases is 3D object detection, which plays a pivotal role in a wide range of applications such as robotics and autonomous driving that require accurate perception of their surrounding environment to operate well \cite{arnold2019survey, qian20223d}.
3D object detection is a fundamental basis in such perception systems, and significant research effort has been devoted to improving its accuracy \cite{lang2019pointpillars, yan2018second, shi2019pointrcnn, shi2020pv, wang2019frustum, simony2018complex}.
It takes 3D sensory data represented by point cloud as input, which is generally captured by \lidar, to generate 3D bounding boxes.
An example is illustrated in Fig.~\ref{fig:2d_vs_3d}.
Compared to its 2D counterpart, 3D object detection introduces a third dimension to characterize the location and size of an object in the real world.
Due to the task complexity and large amount of data to process, 3D detection models usually have more complicated structures with inflated model sizes, posing a higher demand for computation resources.
We experimentally find that with the same hardware, the inference latency of 3D detection model can be up to 41$\times$ of the 2D model (\cref{sec:moti_scenarios}).
Meanwhile, edge computing is provisioned with limited computation power. For instance, the widely used TX2 \cite{tx2} has significantly fewer CUDA cores (17x less) than desktop-class GPU RTX 2080Ti \cite{vs2080ti}, and even fewer (31x less) than high-end GPU Tesla A100 \cite{vsa100}. Thus, it is notoriously difficult to run 3D detection models on edge for near real-time processing.

To address the tension between constrained computing resources and growing demand, the de-facto standard, cloud-only approach offloads compute-intensive 3D object detection to the cloud for inference. Although computation offloading shifts the heavy burden to the cloud and significantly improves inference latency, the end-to-end latency is still unsatisfactory for practical use due to transmission delay, which accounts for the majority of the latency and bottlenecks the entire pipeline (\cref{sec:moti_scenarios}). Despite compression techniques that can reduce point cloud size, the non-negligible compression overhead still hinders real-time streaming.

To enable 3D object detection on constrained edge, we present \sys, a new framework that addresses the limitations of edge- and cloud-only inference, by using 2D models to accelerate 3D object detection. Rather than relying on heavy DNN-based 3D detectors,
\sys proposes a lightweight transformation method that can run on the edge device, with only a few \textit{anchor frames} offloaded to the server for 3D object detection. \sys's approach is based on a \textit{2D-to-3D transformation} method that constructs 3D bounding boxes using outputs from the 2D model. This transformation allows \sys to generate 3D detection results on board, without running heavy 3D models that are ill-suited for edge computing.

However, creating such a full-fledged system involves two key challenges:
(1) At the frame level, how can \sys accurately and efficiently transform 
2D bounding boxes into 3D ones, thus maximizing the latency benefit from 
2D detection?
(2) Across frames, as the error of transformation accumulates over time,
how can \sys monitor the accuracy drop and decide the offloading timing?

\sys addresses these two challenges by introducing a novel system design.
First, \sys starts by running a fast on-board instance segmentation model to obtain both 2D detections and segmentation masks. To utilize previous 3D detection results for better transformation, a \textit{tracking-based association} component is designed to establish the association of 2D bounding boxes in adjacent frames.
Next, we propose the \textit{2D-to-3D Transformation} component, which is a light-weight geometric method that takes in both 2D results and point cloud to generate 3D bounding boxes efficiently.
Specifically, we transfer semantic information contained in the segmentation masks to point cloud to obtain the point cluster of each potential object. Point filtration is designed to eliminate tainted points in each cluster.
Then we design a sequence of geometric methods to accurately construct 3D bounding boxes by referring to previous detection results.
Finally, as the error of transformation accumulates over time, a \textit{frame offloading scheduler} is proposed to judiciously decide when to offload a new frame to launch 3D detectors for the subsequent transformation to utilize.

We implement \sys on an Jetson TX2 and a server equipped with RTX 2080Ti GPU. 
Evaluation on KITTI \cite{kitti} dataset shows that, compared to existing approaches, \sys achieves up to 91.9\% end-to-end latency reduction with only modest accuracy loss. It can even achieve 10 FPS on TX2, matching the scanning frequency of KITTI's LiDAR for real-time processing. Besides, we observe that \sys reduces power consumption and memory usage by up to 75.7\% and 48.1\%, respectively, which is crucial for wimpy edges that need to save resources for more urgent tasks.

In summary, we make three key contributions:
\begin{itemize}
    \item We investigate the system challenges of deploying 3D detection models on edge or cloud and reveal that both edge-only and cloud-only inference incur severe latency overheads, making them ill-suited for latency-sensitive tasks.
    \item We build \sys, the first system that enables 2D-to-3D transformation for robotics and autonomous driving applications. \sys innovatively utilizes previous detection results to enhance the transformation process. Additionally, we introduce a novel approach for coordinating edge and cloud computation using a frame offloading scheduler.
    \item Our experiments demonstrate that \sys offers a significant latency improvement with only modest accuracy loss against several state-of-the-art 3D detection techniques.
\end{itemize}

\section{Background and Motivation}
\label{sec:motivation}

\subsection{The Need for 3D Object Detection}
\label{sec:3d_intro}

State-of-the-art 2D detection techniques lack the depth information to localize objects and estimate their sizes,
which is essential for tasks such as path planning and collision avoidance in
robotics and autonomous driving \cite{arnold2019survey, qian20223d}. 
To overcome these limitations, 3D object detection methods are proposed.
Generally, 3D object detection falls into one of three approaches based on
the input modality \cite{arnold2019survey}. 
The first approach uses the point cloud data from LiDAR or Radar sensors to generate 3D bounding boxes \cite{lang2019pointpillars,
shi2019pointrcnn, yan2018second, shi2020pv, simony2018complex,
zhou2018voxelnet}. 
The second one is image-based 
\cite{reading2021categorical, chen2016monocular, mousavian20173d,
wang2019pseudo, you2019pseudo} that uses neural networks to recover the depth
information from image planes for 3D bounding boxes, which can result in accuracy gaps compared to the point cloud approach.
The third approach is fusion-based \cite{wang2019frustum, qi2018frustum, chen2022focal, 
chen2017multi, vora2020pointpainting, pang2020clocs} which uses one neural network to take both image and point cloud as inputs to improve performance.
We emphasize that although \sys and fusion-based methods both utilize point cloud and images, the key difference is that fusion-based methods use one DNN to extract features from two modalities, while \sys replaces the 3D detection model with the lightweight 2D detection model for most \lidar frames by leveraging an efficient and accurate 2D-to-3D transformation.


\subsection{Challenges of 3D Detection on the Edge}
\label{sec:moti_scenarios}
Despite advancements in hardwares and deep neural networks \cite{arnold2019survey, qian20223d}, performing 3D detection is still a daunting task for resource-constrained edge devices \cite{bhardwaj2022ekya}.
Meanwhile, frequent detection is required to track moving objects, necessitating short detection/inference time on the device.
To clearly illustrate the challenges, 
we first measure the inference latency of 3D object
detection models on a typical edge device, NVIDIA Jetson TX2 \cite{tx2}, with 
one 256-core Pascal GPU. 
Next, to experiment with offloading the 3D object detection models to the cloud, we build a testbed to run the models using a GPU server and replay the real-world network traces to emulate the wide-area network conditions.
Table~\ref{table:pc_based_models} summarizes the four representative point cloud-based models we measure.
The point cloud data comes from the KITTI dataset \cite{kitti}, the most popular one in the field of autonomous driving.


\begin{table}[t]
	\centering
	\small
	\resizebox{0.8\columnwidth}{!}{
	\renewcommand{\arraystretch}{1}
		\begin{tabular}{@{}c|cccc@{}}
			\toprule[0.8pt]
			Model & PointPillar & SECOND & PointRCNN & PV-RCNN \\
			\midrule
			\rowcolor{gray!20}
			\Gape[0pt][2pt]{\makecell[c]{Feature\\Extraction}} & \Gape[0pt][2pt]{\makecell{Voxel\\based}} & \Gape[0pt][2pt]{\makecell[c]{Voxel\\based}} & \Gape[0pt][2pt]{\makecell[c]{Point\\based}} & \makecell[c]{Point-voxel\\based}  \\
			\makecell[c]{Network\\Architecture} & One Stage & One Stage & Two Stages & Two Stages \\ 
			\bottomrule[0.8pt]
		\end{tabular}
	}
	\vspace{2mm}
	\captionof{table}{Point cloud-based models we measure. 
	PointPillar \cite{lang2019pointpillars} and SECOND \cite{yan2018second} group points into vertical columns or voxels and apply convolution over these structures; PointRCNN \cite{shi2019pointrcnn} extracts features directly from the points with permutation invariant learning; PV-RCNN \cite{shi2020pv} combines the point and voxel features.
	}
	\label{table:pc_based_models}
	\vspace{-10mm}
\end{table}

\begin{figure}[t]
	\subfigure[3D models]{\includegraphics[width=0.2\textwidth]{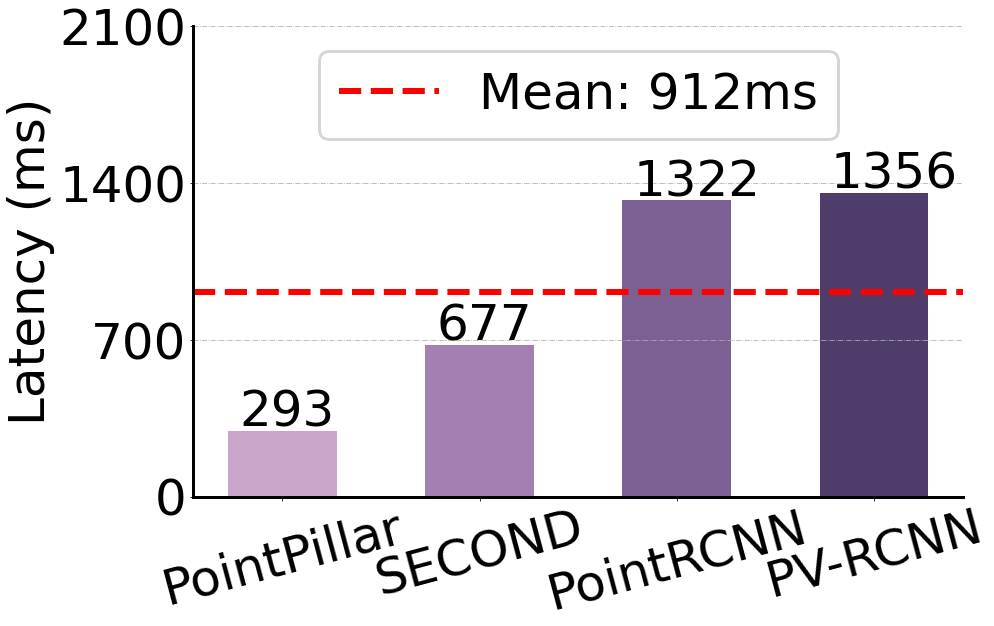}\label{fig:moti_on_board}}\subfigure[2D models]{\includegraphics[width=0.2\textwidth]{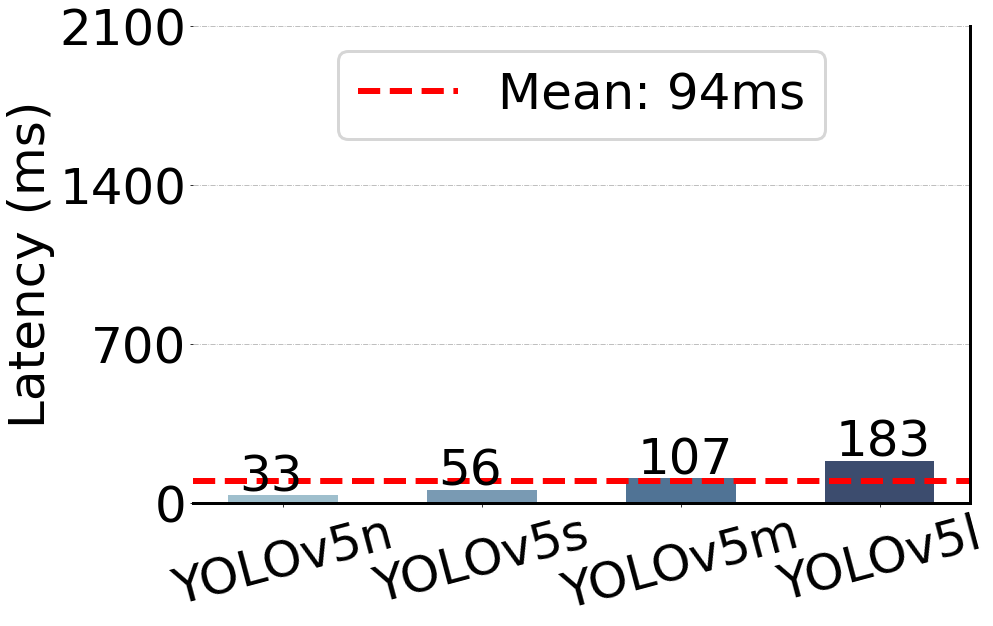}\label{fig:moti_yolo_on_board}}
	\vspace{-6mm}
	\caption{The end-to-end latency of edge-only inference.}
	\label{fig:edge_only}
	\vspace{-5mm}
\end{figure}


\noindent\textbf{Edge-only inference.}
The results, shown in Fig.~\ref{fig:moti_on_board}, indicate that: (i) The average inference latency on the device reaches 912ms for all models, which is impractical for real-world applications that require prompt environmental response; (ii) Even the one-stage models (PointPillar and SECOND) that classify bounding boxes in a single step without pre-generated region proposals have an average inference latency of 293ms and 677ms, respectively; it is not surprising that two-stage approaches take longer due to their more complex pipelines.

To gain a better understanding of the high computational cost of 3D models, we compare them against 2D detection models.
We use four YOLOv5 \cite{yolov5} variants.
\footnote{Here we use YOLOv5's instance segmentation models, which output detection results and segmentation masks simultaneously, and are consistent with our design.}
, and the input images also come from KITTI \cite{kitti}, captured by its car camera with a resolution of 1242$\times$375. 
Results in Fig.~\ref{fig:moti_yolo_on_board} shows that the on-device inference latency of 2D models is much shorter. 
Notably, YOLOv5n only takes 33ms on average; even for the largest YOLOv5l, the inference latency is only 62\% of the fastest 3D model PointPillar. 

\noindent\textbf{Cloud-only inference.}
Given the high latency of edge-only inference, naturally one considers the possibility of performing the inference tasks in the cloud with much more powerful hardware. 
However, with an average size of 6.96Mb per file, streaming point cloud data to the cloud is bandwidth-intensive and offsets the inference latency savings.
To estimate the end-to-end latency of cloud-only inference, we simulate network conditions using bandwidth traces from empirical 4G/LTE datasets from the FCC \cite{fcc} and Belgium \cite{vanderHooft2016},
which are summarized in Table \ref{table:network_traces}.
The point cloud data is uploaded to the server using one TCP connection, and Linux TC \cite{linuxtc} is used to throttle the link capacity according to the bandwidth traces. 
The server performs 3D detection on an RTX 2080Ti GPU. 
Note that we deliberately select the traces that cover different parts of the bandwidth spectrum, and they are all within a normal 4G/LTE cellular performance ($\sim$10--30Mbps) \cite{lte}. It is likely that the real-world performance of cloud-only inference could be worse than what we report here in certain environments \cite{yi2020eagleeye}.

\begin{table}[t]
	\small
		\centering
		\resizebox{0.9\columnwidth}{!}{
			\renewcommand{\arraystretch}{1.3}
				\begin{tabular}{@{}c|ccccc@{}}
					\toprule[1pt]
					Trace (Mbps) & Mean (± Std) & Range & $P_{25\%}$ & Median & $P_{75\%}$ \\
					\midrule
					\rowcolor{gray!20}
					FCC-1 & 11.89 (± 2.83) & [7.76, 17.76] & 9.09 & 12.08 & 13.42 \\ 
					FCC-2 & 16.69 (± 4.69) & [8.824, 28.157] &13.91 & 16.07 & 19.43  \\ 
					\rowcolor{gray!20}
					Belgium-1 & 23.89 (± 4.93) & [16.02, 33.33] & 19.84 & 23.46 & 27.73  \\ 
					Belgium-2 & 29.60 (± 4.92) & [20.17, 37.345] & 25.18 & 30.761 & 32.76  \\ 
					\bottomrule[1pt]
				\end{tabular}
		}
		\captionof{table}{Statistics of four cellular bandwidth traces.}
		\label{table:network_traces}
		\vspace{-8mm}
\end{table}

\begin{table}[t]
	\centering
		\tiny
		\resizebox{0.7\columnwidth}{!}{
		\renewcommand{\arraystretch}{1}
			\begin{tabular}{@{}c|cccc@{}}
				\toprule[0.6pt]
				Algorithm & gzip & zlib & bzip2 & lzma \\
				\midrule
				\rowcolor{gray!20}
				\Gape[0pt][2pt]{\makecell[c]{Compression Time (ms)}} & \Gape[0pt][2pt]{134} & \Gape[0pt][2pt]{238} & \Gape[0pt][2pt]{1007} & \makecell[c]{1179}  \\
				\makecell[c]{Compression Ratio} & 1.57 & 1.57 & 1.75 & 1.83 \\ 
				\bottomrule[0.6pt]
			\end{tabular}
		}
		\captionof{table}{
		The latency and compression ratio of four common compression algorithms.
		}
		\label{table:compression}
		\vspace{-10mm}
\end{table}

\begin{figure}[t]
    \centering
    \subfigure[FCC-1 (Avg. 11.89Mbps)]{\includegraphics[width=0.2\textwidth]{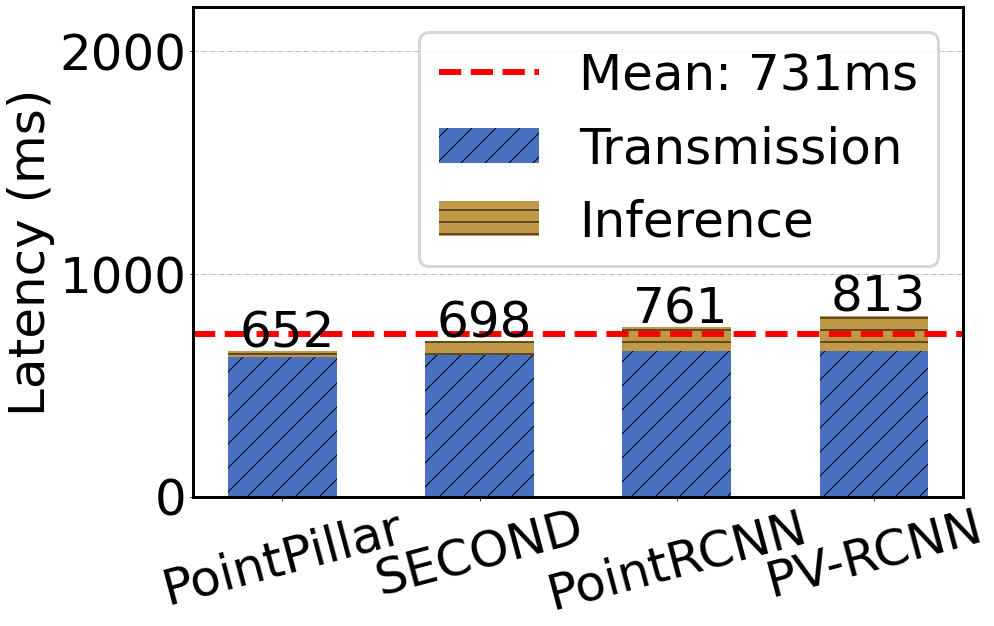}\label{fig:moti_offloading_bw1}}
    \subfigure[FCC-2 (Avg. 16.69Mbps)]{\includegraphics[width=0.2\textwidth]{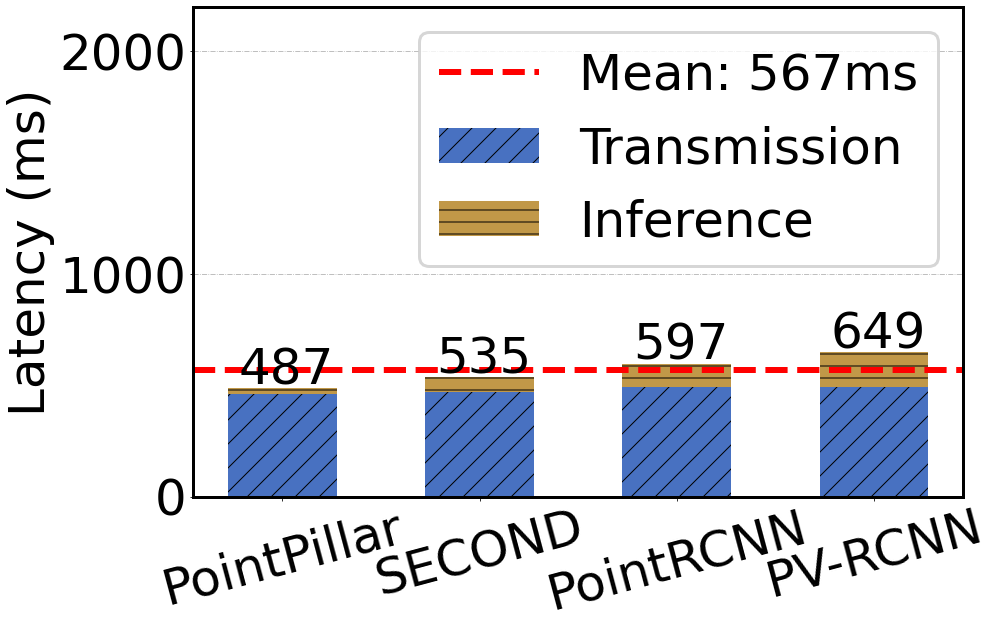}\label{fig:moti_offloading_bw2}} 
    \subfigure[Belgium-1 (Avg. 23.89Mbps)]{\includegraphics[width=0.2\textwidth]{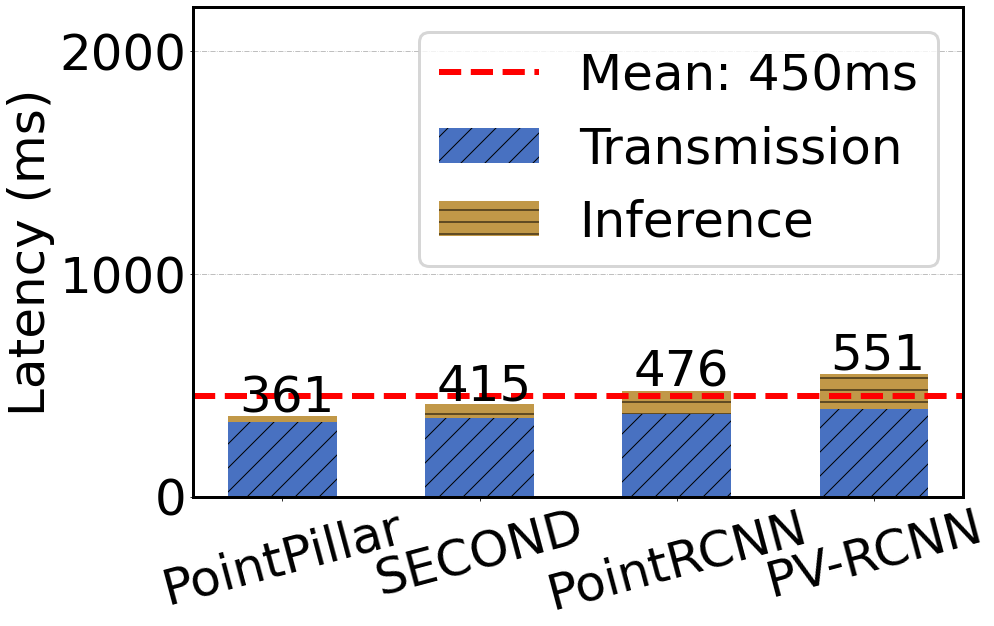}\label{fig:moti_offloading_bw3}}
	\subfigure[Belgium-2 (Avg. 28.60Mbps)]{\includegraphics[width=0.2\textwidth]{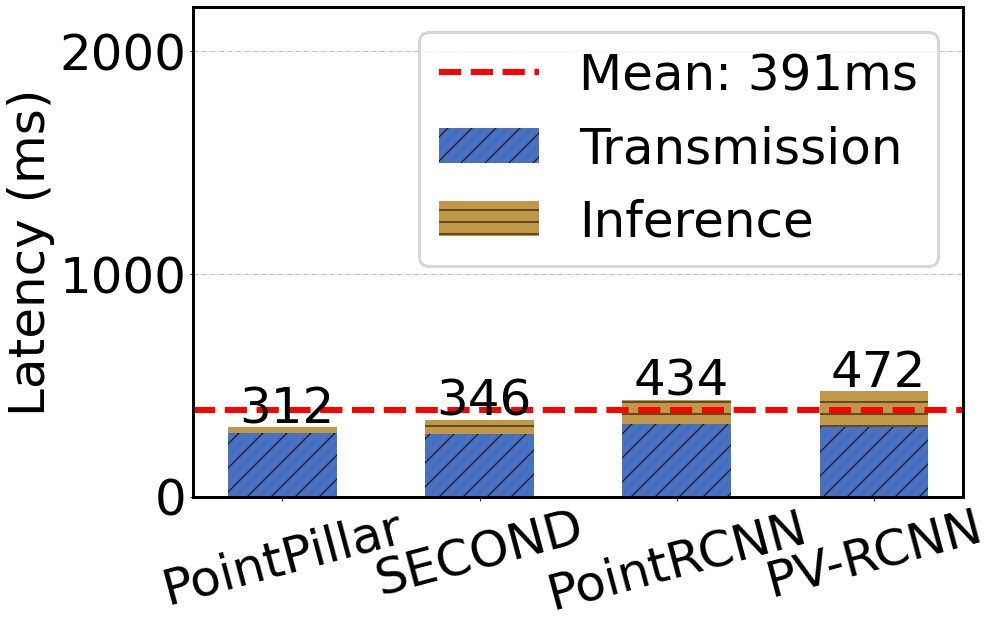}\label{fig:moti_offloading_bw4}}
    \vspace{-4mm}
    \caption{The end-to-end latency of cloud-ony inference on four real-world network traces.}

    \label{fig:moti_offloading}
    \vspace{-6mm}
\end{figure}

As shown in Fig.~\ref{fig:moti_offloading}, we observe that: 
(1) Even for trace with the highest average bandwidth, Belgium 2, the average end-to-end latency across four models reaches 391ms. This is much better than edge-only latency with 912ms mean latency, but it is still not satisfactory for practical use.
(2) The transmission delay over the wide Internet accounts for the majority of latency. When the network condition deteriorates, the end-to-end latency grows noticeably. For instance, the inference time with FCC-1 trace is almost twice that of Belgium-2.
In addition, we also consider compressing the point cloud data before transmission. 
We test the performance of four representative compression algorithms: gzip \cite{gzip}, zlib \cite{zlib}, bz2 \cite{bz2} and lzma \cite{lzma} on TX2, whose CPU runs at 2GHz. We use the implementation in Python standard library to run these algorithms on 50 point cloud files and report the average.
The results are presented in Table.~\ref{table:compression}.
We observe that the compression time is all above 100ms, and a larger compression ratio results in a longer time.
With the fastest algorithim gzip, it can reduce the end-to-end latency of the slowest FCC-1 by 78ms. However, for higher bandwidth traces, compression does not contribute to latency reduction much and can even jeopardize it.
Based on these results, it is evident that offloading all 3D frames to the cloud for inference is also impractical. 

\subsection{Using 2D Models for 3D Detection}
\label{sec:key_idea}

\label{sec:moti_third}
To sum up, the bottleneck of 3D object detection lies in the sheer amount of data that either needs to be processed by complex models on the wimpy edge, or to be transmitted over cellular networks to the cloud. 
Motivated by the drastically lower inference time of 2D object detection (recall Fig.~\ref{fig:edge_only}) and the close correspondence between the 2D and 3D bounding boxes (recall Fig.~\ref{fig:2d_vs_3d}), we cannot help but wonder: what if we use 2D detection models to extrapolate the 3D bounding boxes?
Evidently, this approach would require DNN-based 3D detection on an anchor frame to provide information about the third dimension, and 2D detection can be sufficient to effectively infer the 3D results in subsequent frames.
Additionally, previous detection results of anchor frame can be incorporated to perform better transformation.
To our best knowledge, these have not been explored before. 
\section{System Design}
\label{sec:design}


\subsection{System Overview}

\begin{figure}[t]
	\centering
    \includegraphics[width=0.5\textwidth]{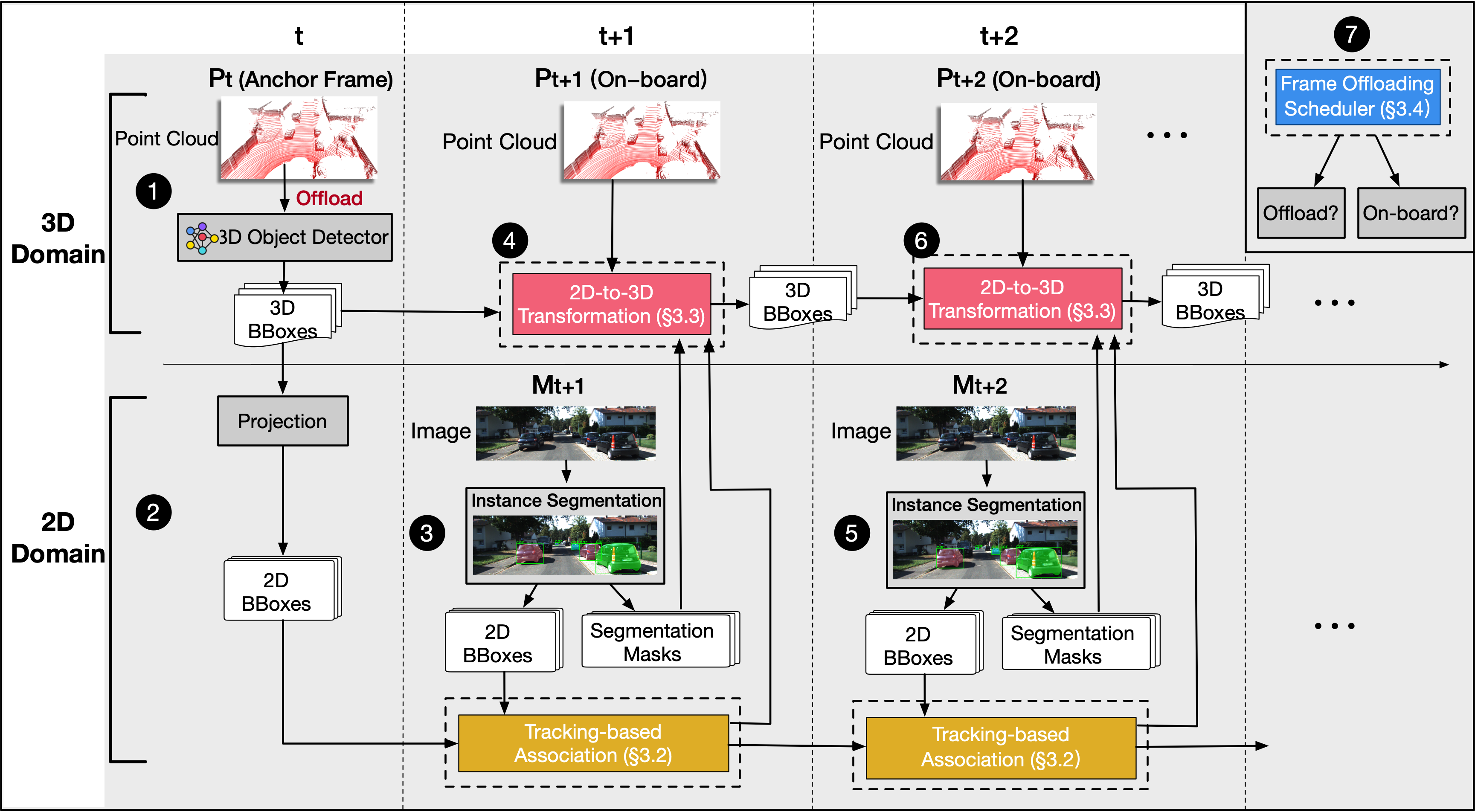}
    \caption{\sys system overview.}
    \label{fig:overview}
    \vspace{-3mm}
\end{figure}

Fig.~\ref{fig:overview} illustrates the overall design of \sys. 
We summarize its workflow into three stages as highlighted in the figure:
\textbf{(1) Preparation}. 
At time $t$, the edge device offloads the \lidar frame $P_t$ (point cloud), i.e. an anchor frame, to the cloud for 3D object detection, and obtains the 3D bounding boxes to initiate the process. The 3D bounding boxes are then projected to 2D ones on the image plane (steps 1\&2 in Fig.~\ref{fig:overview}).
\textbf{(2) Transformation}. 
In the next time slot $t+1$, the edge has both a new \lidar frame $P_{t+1}$ and image $M_{t+1}$ from the camera. 
Instead of running 3D detection on $P_{t+1}$ again, \sys runs an instance segmentation model on $M_{t+1}$ and obtains the 2D bounding boxes and segmentation masks. 
The 2D bounding boxes of both $M_{t+1}$ and $M_t$ are fed into the \textit{tracking-based association} module to build a mapping between the same objects in these two image frames, which serves as the basis for associating the current objects with previous detection results.
Then, using 3D bounding boxes from $P_t$ as the reference, the \textit{2D-to-3D transformation} module takes the segmentation masks and point cloud as input to generate 3D bounding boxes on $P_{t+1}$ (steps 3\&4).
The processing at $t+2$ is identical by using detection mapping and 3D bounding boxes from $t+1$ as the reference (steps 5\&6).
\textbf{(3) Scheduling}. 
\sys's 2D-empowered 3D detection inevitably causes accuracy drop especially as time goes. Thus it relies on a \textit{scheduler} to efficiently monitor the quality of 2D-to-3D transformation and judiciously decide when to offload a new anchor frame to the cloud, so the subsequent transformations have the latest 3D information to draw upon (step 7). 

We now introduce these components in detail.

\subsection{Tracking-based Association}
\label{sec:tracking}
\begin{figure}[t]
	\centering
    \includegraphics[width=0.5\textwidth]{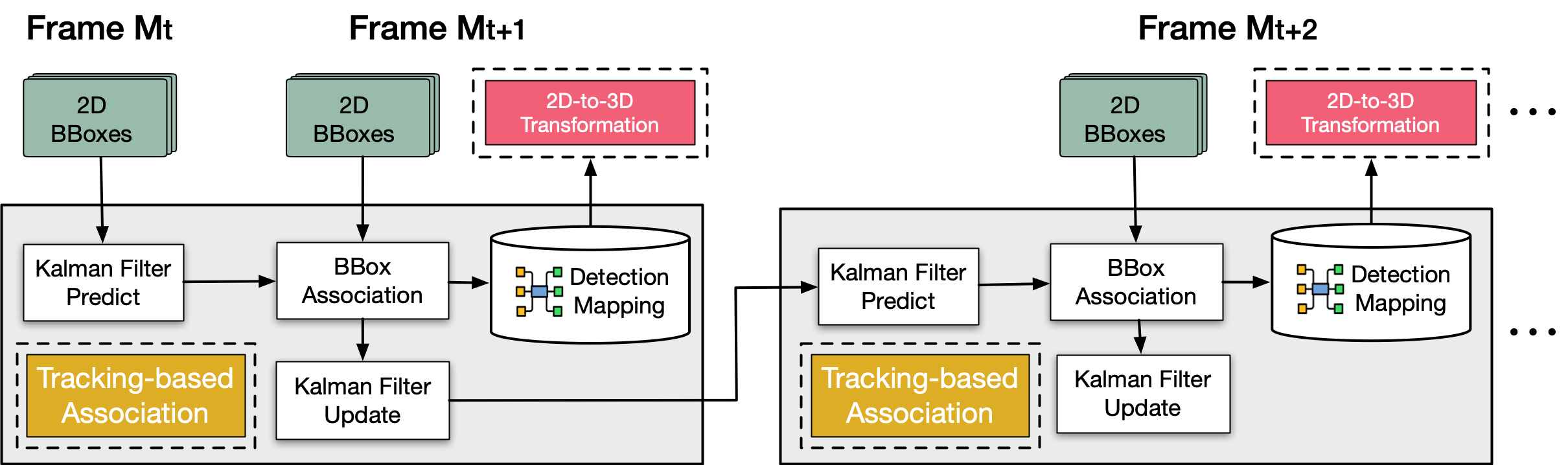}
    \caption{Overview of tracking-based association.}
    \label{fig:tracking}
    \vspace{-3mm}
\end{figure}

The goal of tracking-based association is to utilize tracking in the 2D domain to build the mappings between bounding boxes of the same object in two adjacent frames, which serves as a key basis for 2D-to-3D transformation.
Fig.~\ref{fig:tracking} shows an overview of tracking-based association.

\noindent\textbf{On-device 2D Inference.}
To obtain 2D bounding boxes, Moby either directly projects 3D results if the current \lidar frame is the anchor frame, or runs a 2D detection model. Instance segmentation models are chosen as they output bounding boxes and segmentation masks simultaneously, with bounding boxes used in tracking and segmentation masks needed for the 2D-to-3D transformation.

\noindent\textbf{Kalman Filter-based Tracking.}
\label{sec:kf_based_tracking}
Considering the limited computation resources, the tracking module must ensure real-time processing on edge devices and accurate inter-frame tracking.
We extensively explore existing object tracking techniques in pixel domain and adopt Kalman filter-based tracking \cite{sort}, which satisfies both requirements.
With bounding boxes of each frame, the tracking pipeline (shown in Fig. 5) uses the Kalman filter to predict trajectories from frame $M_t$ to $M_{t+1}$ and estimate the position of boxes in the next frame. Predicted boxes are associated with 2D detection results on $M_{t+1}$ using the Hungarian algorithm \cite{hungarian} and Intersection-over-Union (IoU) criterion, with association rejected if the IoU is below an threshold. The Kalman filter then updates trajectory predictions using matched detections.


\subsection{2D-to-3D Transformation}
\label{sec:transformation}
\begin{figure}[ht]
	\centering
    \includegraphics[width=0.5\textwidth]{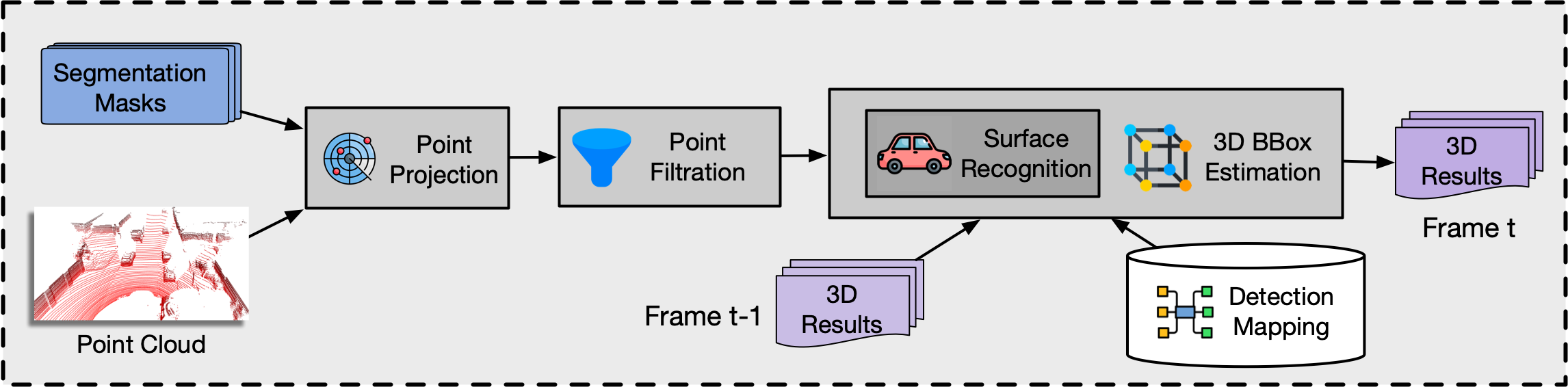}
    \caption{Overview of 2D-to-3D transformation.}
    \label{fig:2d_transformation}
    \vspace{-3mm}
\end{figure}

We now discuss the 2D-to-3D transformation of bounding boxes, which is one of our main technical contributions.
Fig.~\ref{fig:2d_transformation} shows an overview of the 2D-to-3D transformation.
To incorporate 2D semantic information into 3D, \sys first projects the point cloud to the 2D segmentation masks of the image frame in the same time slot.
The point cluster of each object can be identified in 3D;
points of the background that are erroneously labeled as objects are filtered out.
\sys then uses a light-weight geometric method, leveraging previous detection results as reference, to estimate 3D bounding boxes based on point clusters without heavy 3D models.

\begin{figure}[t]
	\centering
    \includegraphics[width=0.5\textwidth]{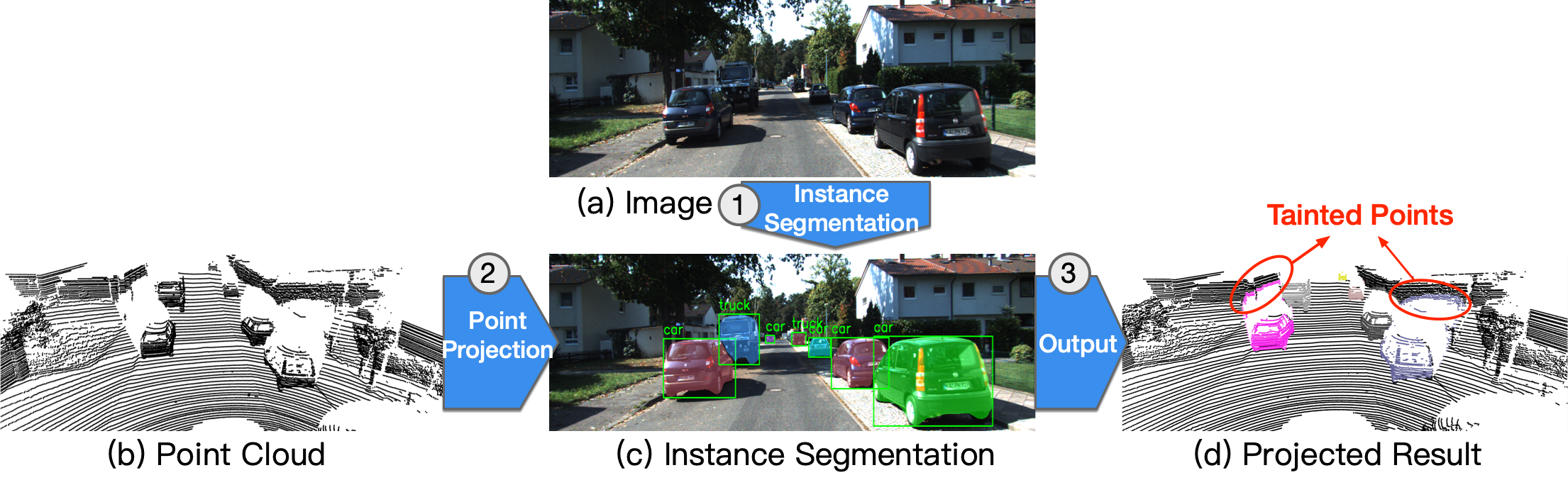}
    \caption{Example of point projection on the KITTI dataset.}
    \label{fig:points_projection}
    \vspace{-3mm}
\end{figure}

\noindent\textbf{Point Projection.}
As discussed in \cref{sec:moti_third}, the 3D and 2D frames captured at the same time have closely related semantic information.
To transfer the semantic information contained in the segmentation masks to 3D, we naturally choose to project the \lidar frame to 2D domain. 
This projection is time-invariant since the camera and \lidar are both fixed once they are installed, and the multimodal-sensor systems today usually provide the transformation matrices in the sensor calibration file to project points from \lidar to camera coordinate.
With the point projection, we can then mark each 3D point to indicate which object it belongs to based on the segmentation masks, by squeezing the stacked masks along the channel dimension.
Then the point cluster of each potential object can be identified.
Fig.~\ref{fig:points_projection} illustrates the process and Fig.~\ref{fig:points_projection}(d) shows the result with point clusters highlighted in different colors.

\noindent\textbf{Point Filtration.}
Directly transferring segmentation masks to the point cloud may cause certain points to be erroneously marked as objects of interest and degrade accuracy. 
This can also be seen in Fig.~\ref{fig:points_projection}(d), where some background points are recognized as part of a vehicle since they are projected to the same region of the vehicle in 2D \cite{qian20223d}. 

\begin{figure}[t]
	\centering
    \includegraphics[width=0.5\textwidth]{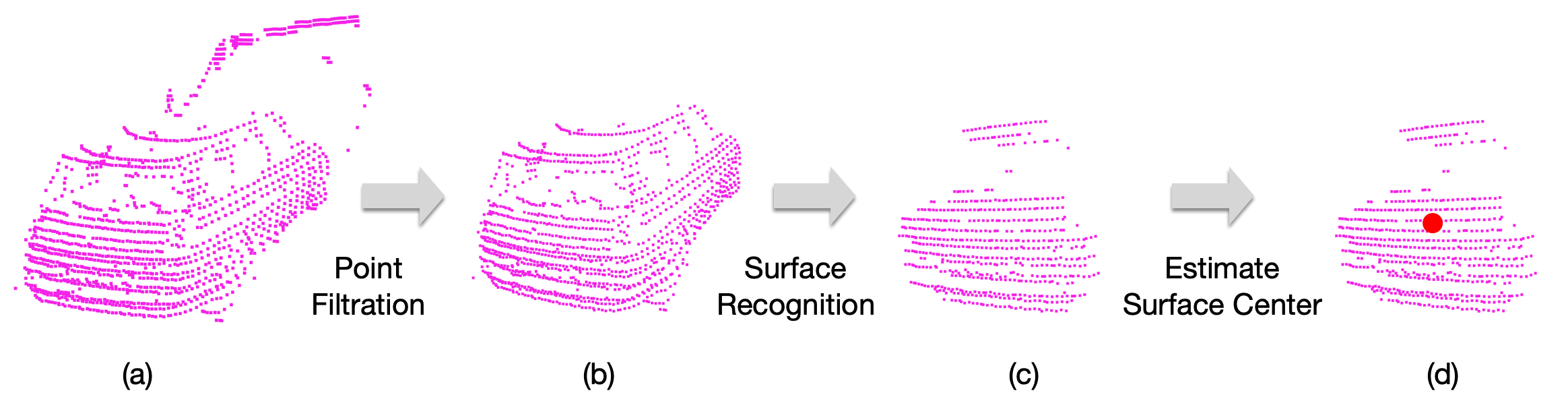}
    \vspace{-6mm}
    \caption{Key steps of point filtration and 3D bounding box estimation for a point cluster.}
    \label{fig:plane_fitting}
    \vspace{-5mm}
\end{figure}

To filter out these tainted points, we design an algorithm to ``purify'' each object's point cluster for better 3D bounding box estimation. 
The details are summarized in Algorithm~\ref{algo:points_filtration}.  
Chiefly, for each point cluster, it first calculates the distance from all points to the origin of \lidar coordinate (line \ref{alg:1}).
Next, it searches for the nearest point to the origin, which most likely represents the boundary of this object and is thus called the \textit{critical boundary point} (line \ref{alg:3}).
Then it calculates the distance from all points to the critical boundary point (line \ref{alg:4}).
Those points close to the critical boundary point are considered to belong to the object itself (line \ref{alg:5}).
If too few points are filtered this way, it suggests that the critical boundary point may not be the actual boundary of this object, which can happen when for instance vehicles are close to each other. 
We then add a small step size $S_T$ and use the nearest point whose distance to the origin is at least $S_T$ further away as the new critical point (line \ref{alg:2}).
We repeat the process until it finds a cluster that has enough points or the number of iteration exceeds three (lines \ref{alg:7} and \ref{alg:6}).
Fig.~\ref{fig:plane_fitting} depicts an example of the effect of point filtration. The rationale of the algorithm is based on our observation that the points of potential objects are commonly much nearer to the origin than background points.
As we measured quantitatively, our point filtration algorithm removed 98\% of tainted points.
After point filtration, the point cluster of each potential object is clean enough for the following 3D bounding box estimation.

\begin{algorithm}[t]
    \caption{Point Filtration}
    \label{algo:points_filtration}
    \footnotesize

    \SetKwFunction{FMain}{\underline{point\_filtration}}
    \SetKwProg{Fn}{Function}{:}{}
    \SetKwInOut{Input}{Input}\SetKwInOut{Output}{Output}
    \SetKwRepeat{Do}{do}{while}

    \Input{
            
        - $PC_{old}$: \lidar points of all potential objects after point projection.

        - $F_T$: The filtering threshold of Euclidean distance range.

        - $M_T$: The threshold of minimum points in a potential object.

        - $S_T$: The threshold of step size.

    }
    \Output{$PC_{new}$: Filtered \lidar points.
    }
    \Fn{\FMain{$PC_{old}, F_T, M_T, S_T$}}{
        \For{$PC_i$ in $PC_{old}$}{
            $PC_{new}$, $idx$, $iter$ $\leftarrow$ \{ \}, [ ], $0$\;
            $init\_list$ $\leftarrow$ calculate the Euclidean distance from all points to the origin of \lidar coordinate\; \label{alg:1}
            $n_j$ $\leftarrow$ get the nearest point to the origin according to $init\_list$\;\label{alg:3}
            \While{$sum(idx) < M_T$}{\label{alg:7}
                $dist\_list$ $\leftarrow$ calculate the distance from all points to $n_j$\; \label{alg:4}
                $idx$ = $dist\_list$ $<$ $F_T$\; \label{alg:5}
                $n_j$ $\leftarrow$ get the point whose distance to origin is at least $S_T$ further away\; \label{alg:2}
                $iter++$\;
                \lIf{$iter$ == $3$}{\label{alg:6}
                    break
                }
            }
            $PC_i' = PC_i[idx]$\;
            $PC_{new}$ $\leftarrow$ $PC_{new} \bigcup$ $PC_i'$\;
        }
    \Return{$PC_{new}$};}
\end{algorithm}

\noindent\textbf{3D Bounding Box Estimation}
\label{design:box_estimation}
\sys now strives to estimate each object's 3D bounding box based on its point cluster. 
A 3D bounding box is represented by a seven-tuple $[x, y, z, l, w, h, \theta]$, including the object's center $[x, y, z]$, size $[l, w, h]$, and heading angle $\theta$ relative to the $x$ axis on the $x-y$ plane of \lidar coordinates. 
It is challenging to estimate all parameters solely based on the values of points, especially the heading and center, for two main reasons: 
1) The point cloud is sparse and irregular; 2) Only part of the object has a point cloud because of the self-occlusion problem \cite{qian20223d}. 

\begin{figure}[t]
	\begin{minipage}{.6\linewidth}
        \centering
        \includegraphics[width=1\textwidth]{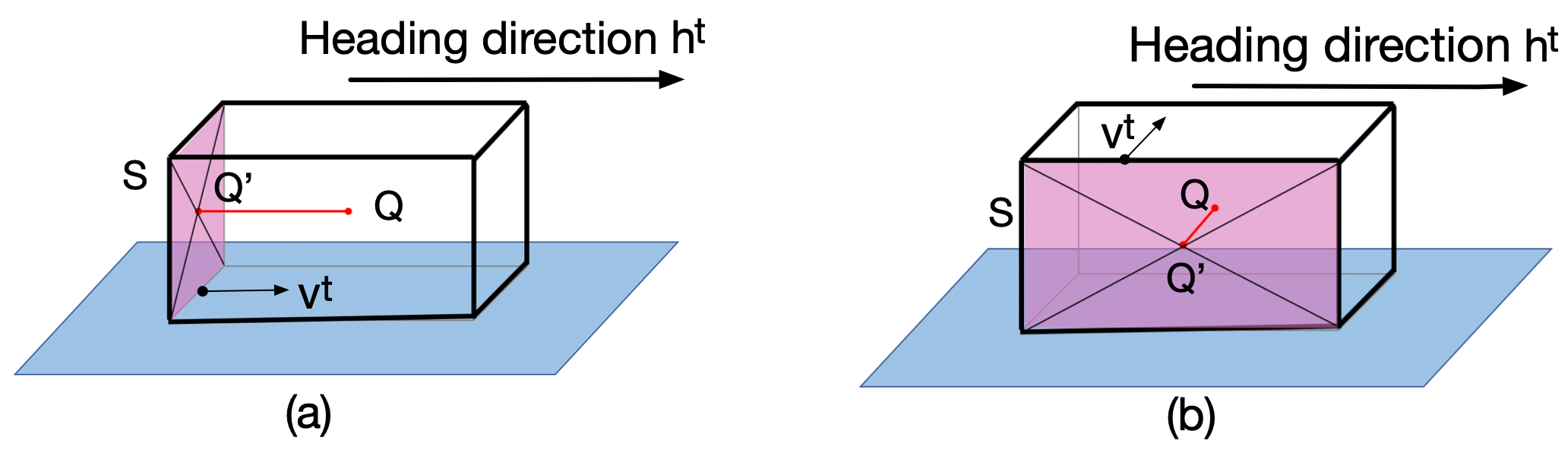}
        \caption{Two possible situations in bounding box estimation. 
        }
        \label{fig:two_situations}
	\end{minipage}
	\begin{minipage}{.38\linewidth}
	  	\centering
		\includegraphics[width=0.495\textwidth]{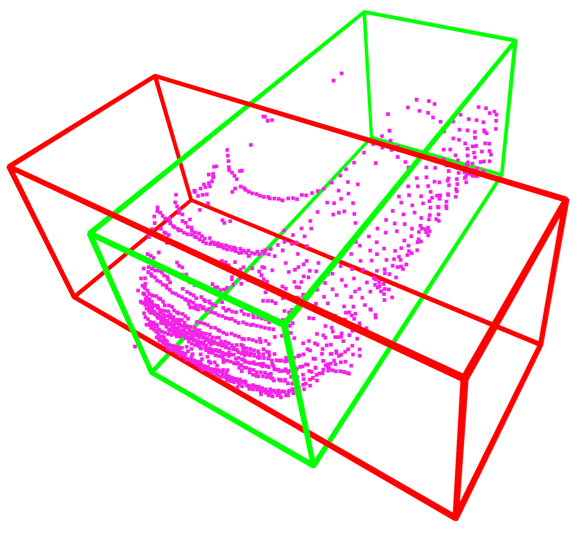}
		\caption{Compare the number of points inside both situations to determine the correct box.}
        \label{fig:two_boxes}
	\end{minipage}
    \vspace{-5mm}
\end{figure}

We introduce a novel approach for accurately and efficiently estimating 3D bounding boxes.
For each point cluster that has been associated with an object from the previous \lidar frame by the association module, \sys first obtains the object's size directly from the previous frame's result (size of the same object does not change). 
It then estimates the heading angle, and calculates the object center based on size and heading.


Now the heading angle can be obtained from the normal vector of one of the object's surfaces, if we know which side this surface is. Fig.~\ref{fig:two_situations} shows an example of this intuition with different surfaces of an object.
Based on this idea, 
\sys uses the well-known RANSAC algorithm \cite{ransac} to find a surface first. It iteratively selects three random points to form a plane until a best-fitting plane is found with the most inliers in it, with its normal vector $v_t$.
Fig.~\ref{fig:plane_fitting}(c) shows the plane it finds on the point filtration output. 
This plane can be either the front or rear surface as in Fig.~\ref{fig:two_situations}(a), or the side surface as in Fig.~\ref{fig:two_situations}(b).\footnote{Theoretically, the found plane can also be the top surface of a vehicle. Yet in the autonomous driving scenario, \lidar are installed at the top of the vehicle, and the angle of laser beams makes it very unlikely for the top surface to have more points and then be found by RANSAC. Even when this happens, it can be handled by removing points in the top surface and re-run RANSAC.}
To determine the correct heading, \sys calculates the angle between the normal vector $v^{t}$ of the surface and the heading direction $h^{t-1}$ of this object's associated bounding box in the previous \lidar frame. 
If the angle is less than a threshold $\xi$ or greater than $\pi - \xi$, it means the normal vector $v^{t}$ is nearly parallel to $h^{t-1}$, and the surface is the rear or front as shown in Fig.~\ref{fig:two_situations}(a).
Further, since the object moves continuously and is physically impossible to change its heading dramatically in one frame (i.e. 0.1s in KITTI), \sys obtains the heading direction $h^{t}$ directly from $v^t$:

\begin{equation}
    \begin{split}
    h^{t} =
    \begin{cases}
        v^{t}, & \text{if angle between }h^{t-1} \text{ and } v^{t} \text{ is less than } \xi, \\
        -v^{t}, & \text{if angle between }h^{t-1} \text{ and } v^{t} \text{ is greater than } \\ & \pi - \xi \text{ ($v^t$ in the opposite direction)}.
    \end{cases}
    \label{equ:heading_angle}
    \end{split}
\end{equation}

In case $v^t$ is perpendicular to $h^{t-1}$ as in Fig.~\ref{fig:two_situations}(b), we can also easily obtain $h^{t}$ by rotating $v^{t}$ by either 90 or 270 degrees. The heading angle $\theta$ is then directly calculated from $h^t$.

Lastly, in terms of the object center $Q$, \sys estimates it based on the surface center $Q'$, the heading angle $\theta$, and the object size.
Note that the surface center $Q'=[a, b, c]$ is obtained by averaging the corresponding value of all points in the plane, as shown in Fig.~\ref{fig:plane_fitting}(d). Then the object center $Q$ can be directly calculated by

\begin{equation}
    Q =
    \begin{cases}
      [a, b, c] + [0.5*l*\cos{\theta}, 0.5*l*\sin{\theta}, 0], \text{if angle between }
      \\\hspace{\parindent}\hspace{\parindent}\hspace{\parindent}\hspace{\parindent} h^{t-1} \text{ and } v^{t} \text{ is less than } \xi \text{ or greater than } \pi - \xi,\\
      [a, b, c] + [0.5*w*\cos{\theta}, 0.5*w*\sin{\theta}, 0], \hspace{\parindent}\hspace{\parindent}\text{otherwise.}
    \end{cases}
    \label{equ:surface_center}
\end{equation}

Finally we need to consider objects that are not associated with anything in the previous frame after tracking-based association.
These are highly likely new objects that first appear in the current frame.
Since we no longer have the reliable size information, we use the average length, width and height of all objects to estimate the new object's size.
This can be updated after \sys schedules another anchor frame to go through the 3D detector. 
To estimate the heading angle, \sys uses the same procedure as described above to identify a best-fitting surface based on the point clusters from point filtration. 
As we no longer have prior knowledge of the heading direction (from the previous \lidar frame), we adopt an alternative method by calculating the bounding boxes for both possibilities by Eq.~\eqref{equ:surface_center}, and the one that fits more points inside is the final output as shown in Fig.~\ref{fig:two_boxes}.

\subsection{Frame Offloading Scheduler}
\label{sec:scheduler}

\begin{figure}[t]
	\begin{minipage}{.53\linewidth}
        \centering
        \includegraphics[width=1\textwidth]{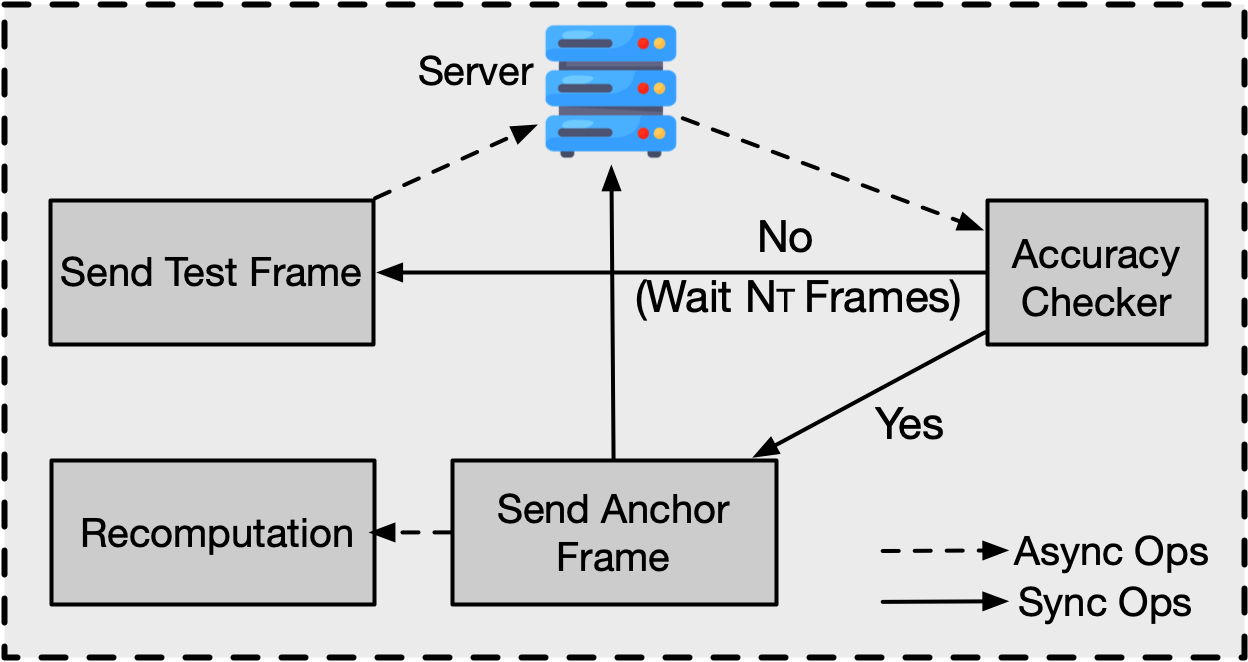}
        \caption{Overview of frame\\ offloading scheduler.}
        \label{fig:frame_scheduler}
	\end{minipage}
	\begin{minipage}{.45\linewidth}
	  	\centering
        \includegraphics[width=1\textwidth]{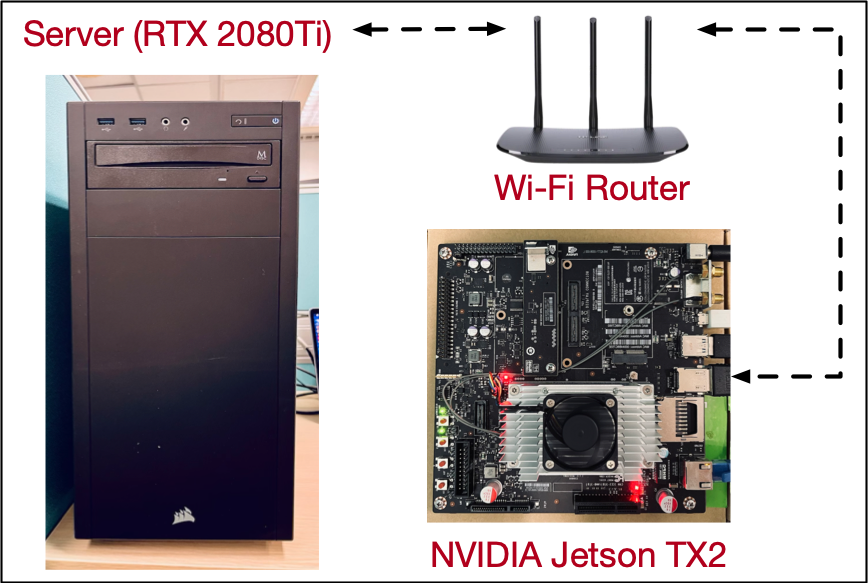}
        \caption{Our evaluation hardware platform.}
        \label{fig:impl}
	\end{minipage}
    \vspace{-6mm}
\end{figure}

\sys's 2D-to-3D transformation is based on the 3D detection results of the precedent anchor frame. As time goes this transformation becomes less effective simply because the scene and the objects may have changed quite much from the anchor frame. 
Thus from time to time, \sys needs to offload a new anchor frame to the cloud for 3D detection to assist the subsequent transformation and maintain overall accuracy.
To efficiently schedule frame offloading, we must be able to determine when a new anchor frame is needed.  
For this purpose, a test \lidar frame is sent to the cloud every $N_T$ frames. Detection on the test frames happen in parallel with \sys's on-device processing.  
We deem the 3D detection results as ground truth to obtain the error of the 2D-to-3D transformation on the same test frame. 
When the error is larger than a threshold, the next \lidar frame is designated as the new anchor frame and also sent to the cloud for inference. The subsequent on-device processing is blocked until results of the new anchor frame is received by the edge device. 
Fig.~\ref{fig:frame_scheduler} depicts the complete process.
Our design strikes a fine balance between overheads and effectiveness.


\noindent\textbf{Recomputation.}
\label{sec:recomputation}
Since the process of waiting anchor frame's result is synchronous, the operational flow of \sys is blocked to wait for the result from server to continue the on-board processing.
To make use of the idle computation resources on edge device during the waiting time, we propose a mechanism called \textit{recomputation} to recompute the 3D results of past intermediate frames using the results of test frames. 
Specifically, some intermediate results such as outputs of instance segmentation model are stacked.
When an anchor frame is triggered, the intermediate results and the stale output of test frame are utilized to re-run the 2D-to-3D transformation to recompute 3D bounding boxes. We observe that the recomputation time is much less than the offloading time and is completely hidden without inducing extra latency.
\section{Implementation}
\label{sec:implementation}

Our framework is implemented with $\sim$4K lines of Python. We use a Jetson TX2 \cite{tx2} as the edge device and a desktop with Intel i7-9700K CPU and RTX 2080Ti GPU as the server.
As shown in Fig.~\ref{fig:impl}, they are physically connected to a Wi-Fi router and communicate via TCP.
We deploy OpenPCDet \cite{openpcdet2020} on the server as the inference engine.
The filtering threshold $F_T$, number of minimum points $M_T$ and step size $S_T$ are empirically set to 4.5, 24 and 12 in point filtration.
$\xi$ in the box estimation is 30 degrees.
As for the offloading scheduler, the accuracy threshold and the gap of test frame, i.e. $Q_T$ and $N_T$, are set to 0.7 and 4, respectively.
\section{Evaluation}
\label{sec:evaluation}
We now evaluate \sys by answering the following questions:
\begin{itemize}[noitemsep,topsep=0pt,parsep=0pt,partopsep=0pt,leftmargin=*]
    \item How well does \sys work in terms of latency and accuracy?
    \item How much does each key design choice contribute to the overall performance gain?
    \item How sensitive is \sys to the key parameters?
    \item How efficiently does \sys utilize the edge resources?
\end{itemize}

\subsection{Experimental Setup}
\label{sec:eval_setup}

\noindent\textbf{Datasets.}
All experiments are done on the KITTI dataset \cite{kitti}, 
a real-world autonomous driving benchmark, with synced \lidar scans and camera images at 10 FPS.
It provides 3D ground truth for cars, pedestrians, and cyclists. 
Currently we only focus on the most challenging class, car, as it moves the fastest.
We use the same empirical bandwidth traces as shown in Table~\ref{table:network_traces}.

\begin{figure*}[ht]
    \centering
    \subfigure[FCC-1 (Avg. 11.89Mbps)]{\includegraphics[width=0.19\textwidth]{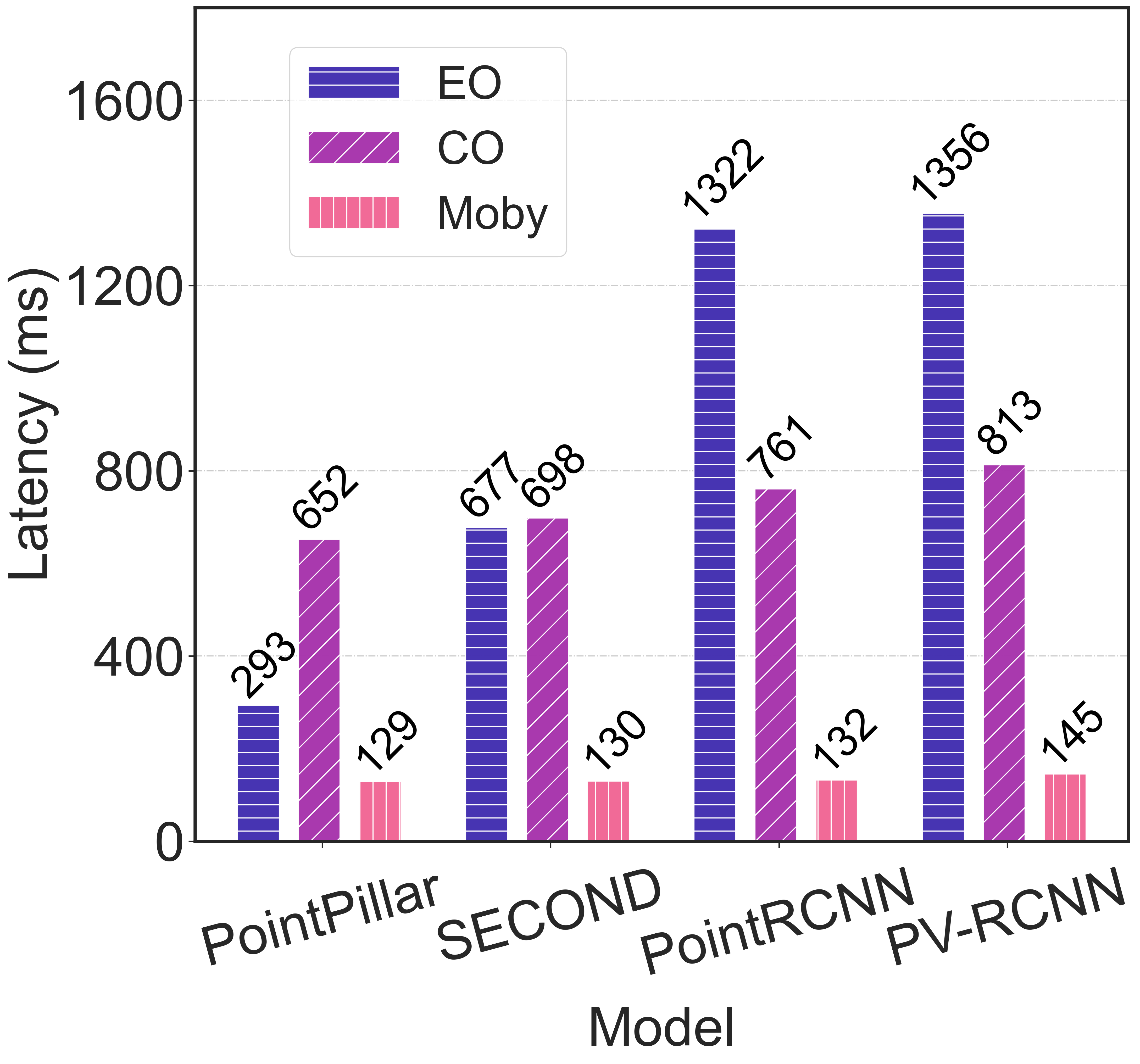}\label{fig:overall_bw4_latency}}
    \subfigure[FCC-2 (Avg. 16.69Mbps)]{\includegraphics[width=0.19\textwidth]{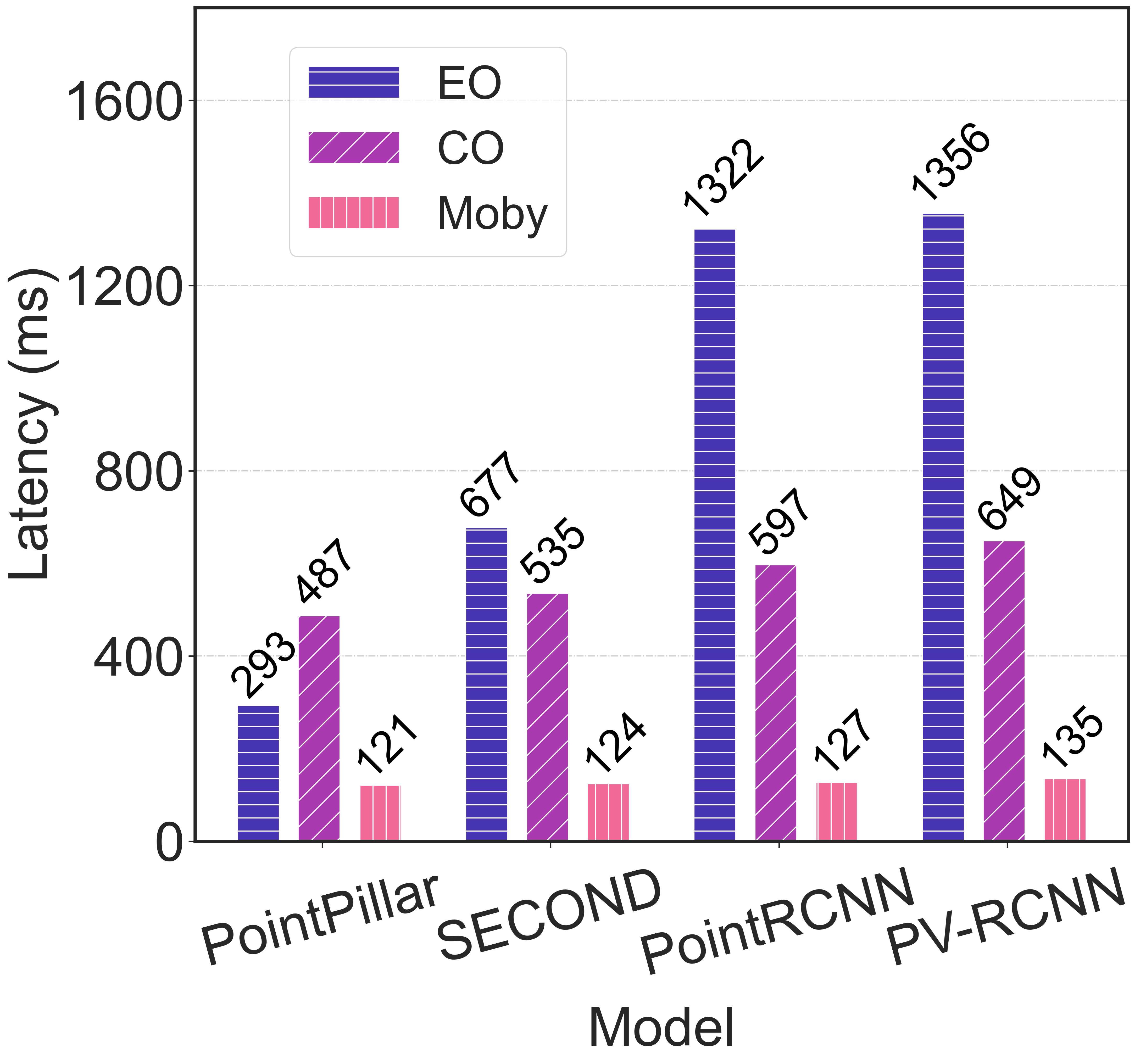}\label{fig:overall_bw3_latency}} 
    \subfigure[Belgium-1 (Avg. 23.89Mbps)]{\includegraphics[width=0.19\textwidth]{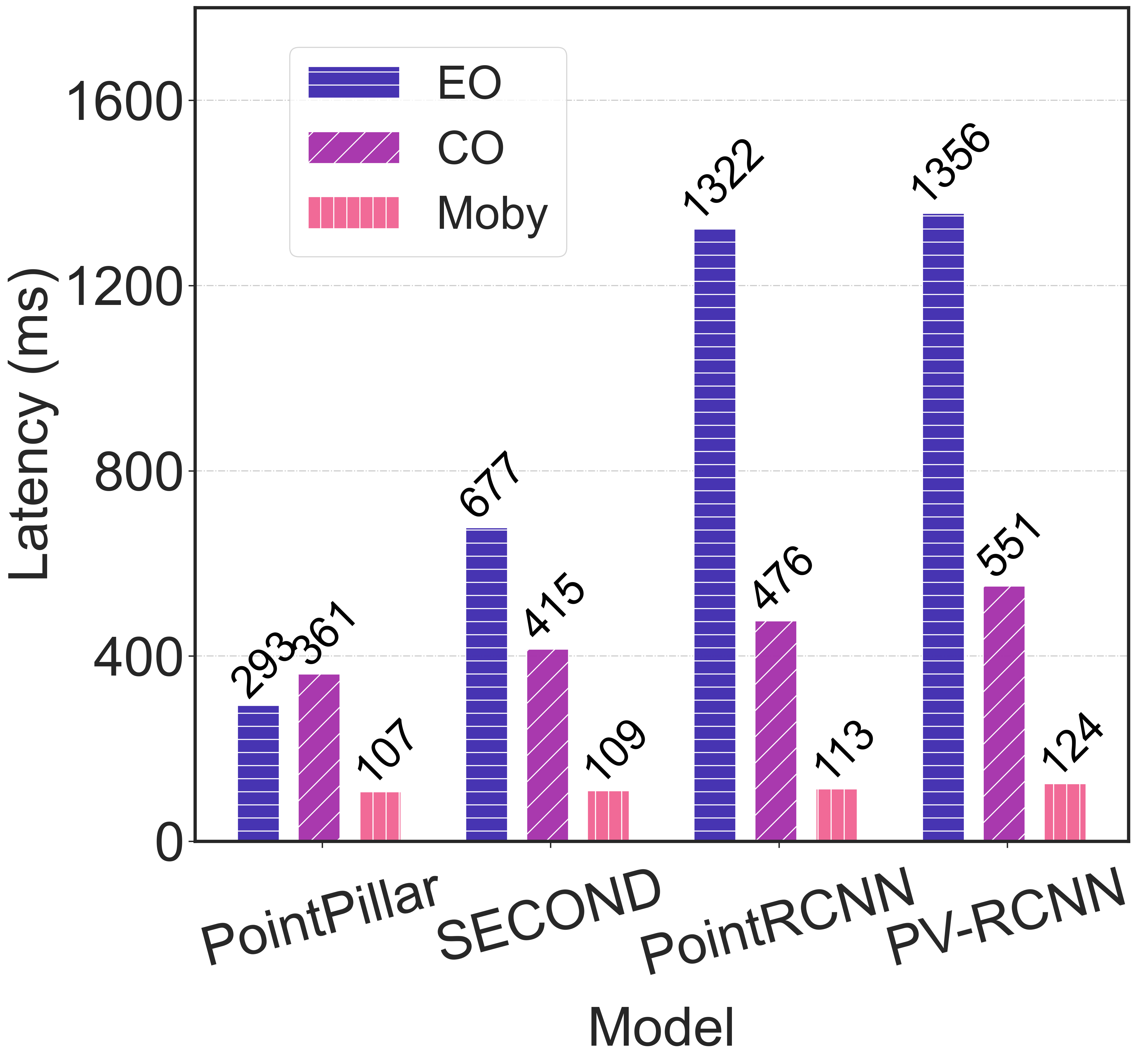}\label{fig:overall_bw2_latency}}
	\subfigure[Belgium-2 (Avg. 28.60Mbps)]{\includegraphics[width=0.19\textwidth]{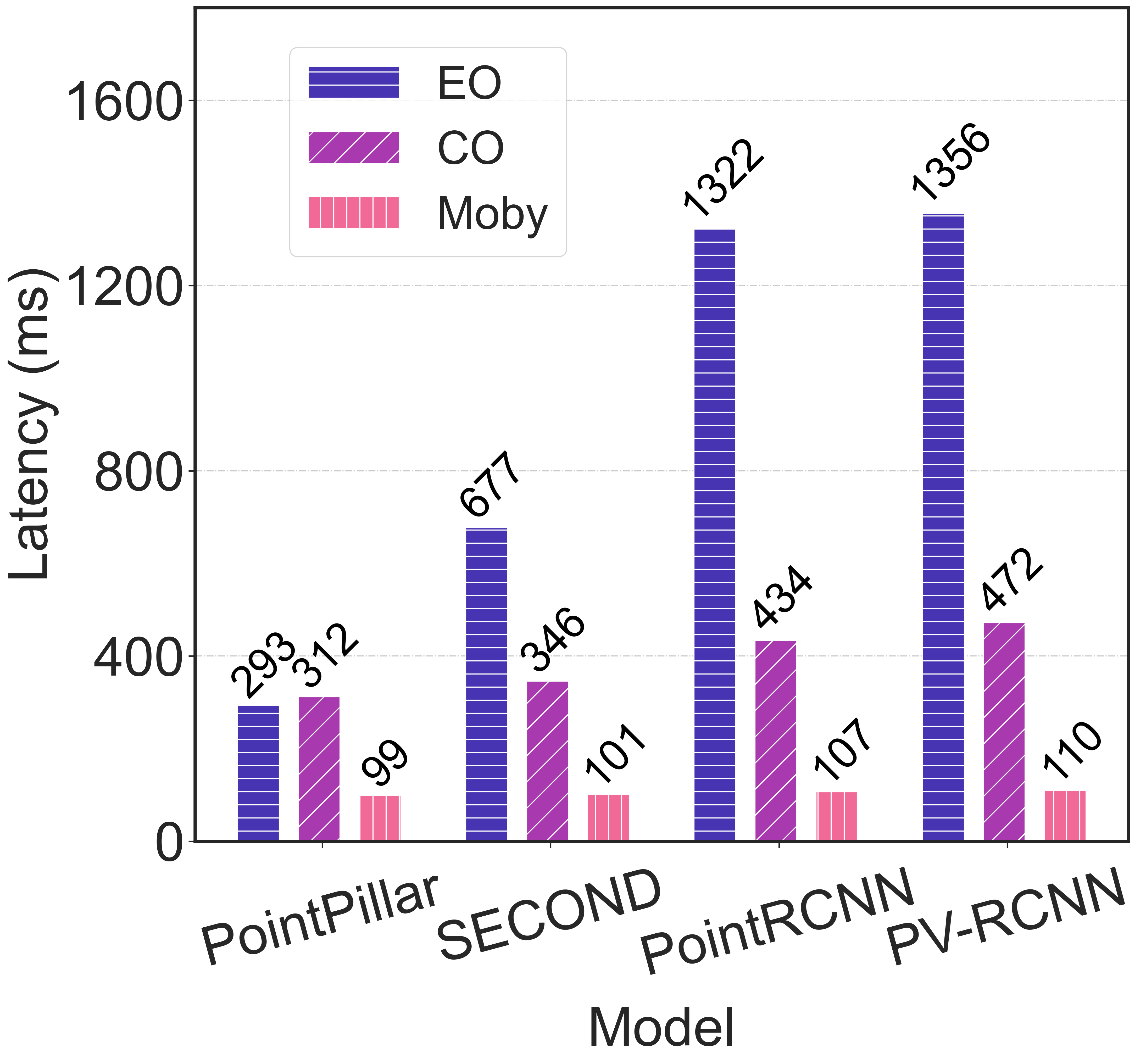}\label{fig:overall_bw1_latency}}
    \subfigure[Average accuracy]{\includegraphics[width=0.19\textwidth]{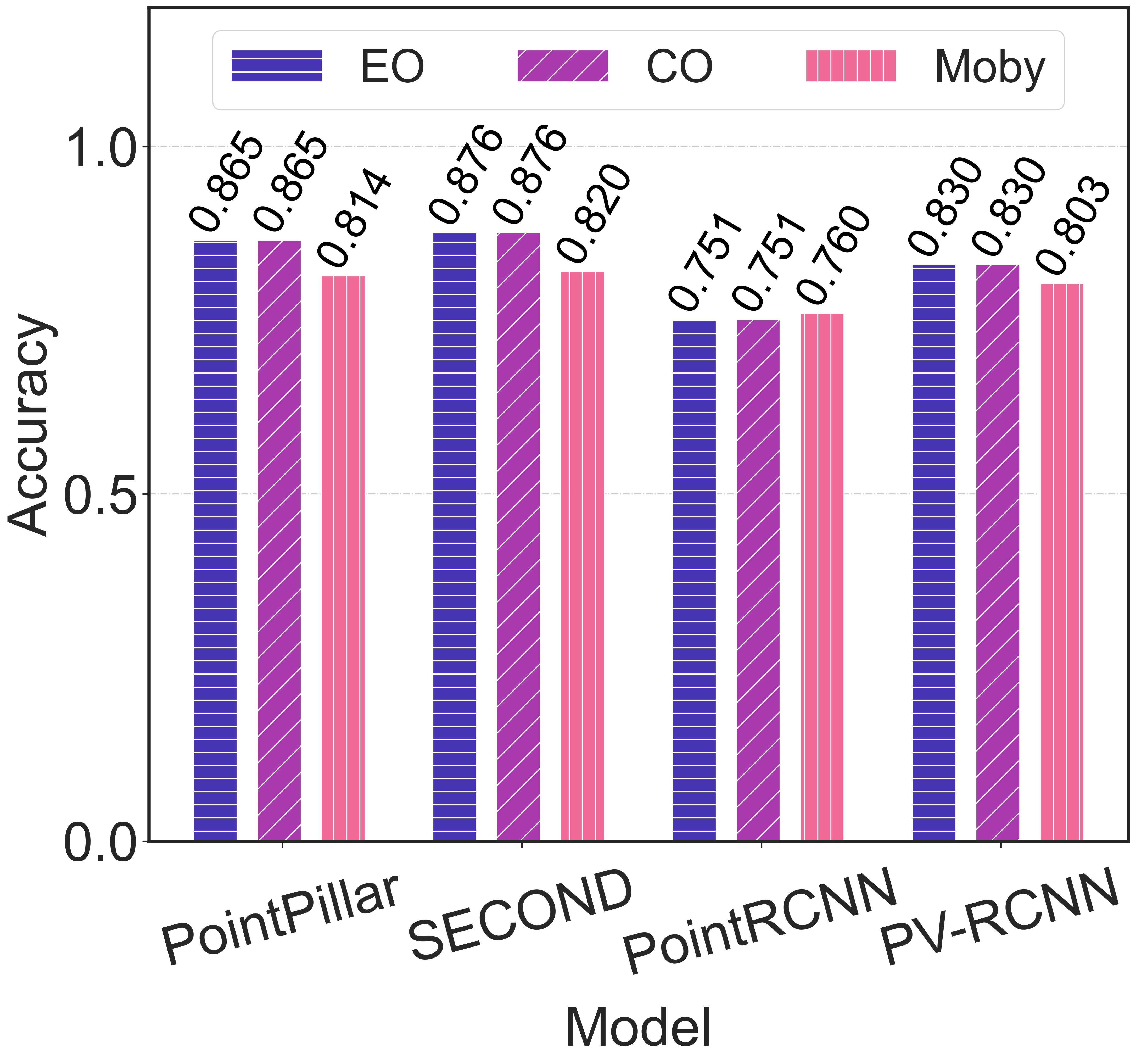}\label{fig:overall_accuracy}}
    \vspace{-4mm}
    \caption{The latency (a-d) and accuracy (e) comparison of \sys, edge-only, and cloud-only deployment approaches.}

    \label{fig:overall_latency}
    \vspace{-4mm}
\end{figure*}

\noindent\textbf{Performance Metrics.}
We use the end-to-end latency and accuracy as main performance metrics.
The end-to-end latency is the time between the input of point cloud cloud and the output of 3D detection results.
We consider the object to be successfully detected if the 3D IoU between detection and ground truth exceeds 40\% (i.e., accurately located).
F1 score is used to measure accuracy, i.e. the harmonic mean of precision and recall, as widely adopted in existing work \cite{shuai2021millieye, du2020server, zhang2018awstream, zhang2017live, chen2015glimpse}.

\noindent\textbf{Models.}
We adopt YOLOv5n from the YOLOv5 codebase\cite{yolov5}
as \sys's default instance segmentation model.
Unless otherwise specified, \sys uses the same 3D object detection model as the baseline systems.
Note that \sys is not model-dependent: any 3D object detection and instance segmentation model can be applied.

\subsection{Overall Performance}
\label{sec:eval_overall}

We evaluate \sys comprehensively from two different perspectives: 
(1) deployment approach. Here we show how well \sys performs compared to two common scenarios of edge inference, i.e., completely running on the edge and offloading to cloud;
(2) acceleration method. Here we compare \sys to other methods that accelerate 3D object detection.

\subsubsection{Deployment Approaches.}
We first compare \sys with the following two common deployment approaches:
\begin{itemize}[noitemsep,topsep=0pt,parsep=0pt,partopsep=0pt,leftmargin=*]
    \item \textbf{\em Edge Only (EO)}: In this scheme, the 3D models are deployed on the edge device only to run inference. 
    
    \item \textbf{\em Cloud Only (CO)}: The point cloud data is fully offloaded over 4G/5G networks to the server for inference. The end-to-end latency involves both transmission and inference delay.

\end{itemize}

Fig.~\ref{fig:overall_latency} shows the latency comparison results using four representative point cloud based 3D detection models, as described in Table~\ref{table:pc_based_models}. 
Observe that \sys outperforms EO and CO in latency with significant margins ranging from 56.0\% to 91.9\% across models.
Notably, \sys achieves the lowest latency 99ms with PointPillar in Belgium-2 trace,
matching the 10 FPS frame rate in KITTI for real-time processing.
The latency gain is much more salient for two-stage models, such as PointRCNN and PV-RCNN, as their architectures pose much severer computation burden.
\sys delivers speedups across bandwidth traces.
Even for the lowest bandwidth trace, FCC-1, \sys reduces latency by 56.0\% compared to running the fastest PointPillar onboard. As the bandwidth increases, the latency gain reaches 66.2\%.
The latency improvement mainly comes from the design choice of replacing the 3D object detector with a much cheaper 2D model that can run on the edge device, which dramatically reduces both the inference time and transmission time (to the cloud).


\sys has little impact on 3D object detection accuracy. Fig.~\ref{fig:overall_accuracy} reports overall accuracy for all schemes.
Note that as EO and CO both run 3D object detectors for all frames, they have the same accuracy with the same model. 
For PointRCNN, \sys achieves almost the same or slightly higher accuracy (0.760 vs 0.751) due to the effective 2D-to-3D transformation mechanism. For the other three models, accuracy drops slightly between 0.027 to 0.056 due to: (i) possible occlusion problem in 2D images and (ii) bounding box estimation errors caused by sparse point clusters.

\begin{figure}[t]
    \centering
    \subfigure[Latency]{\includegraphics[width=0.21\textwidth]{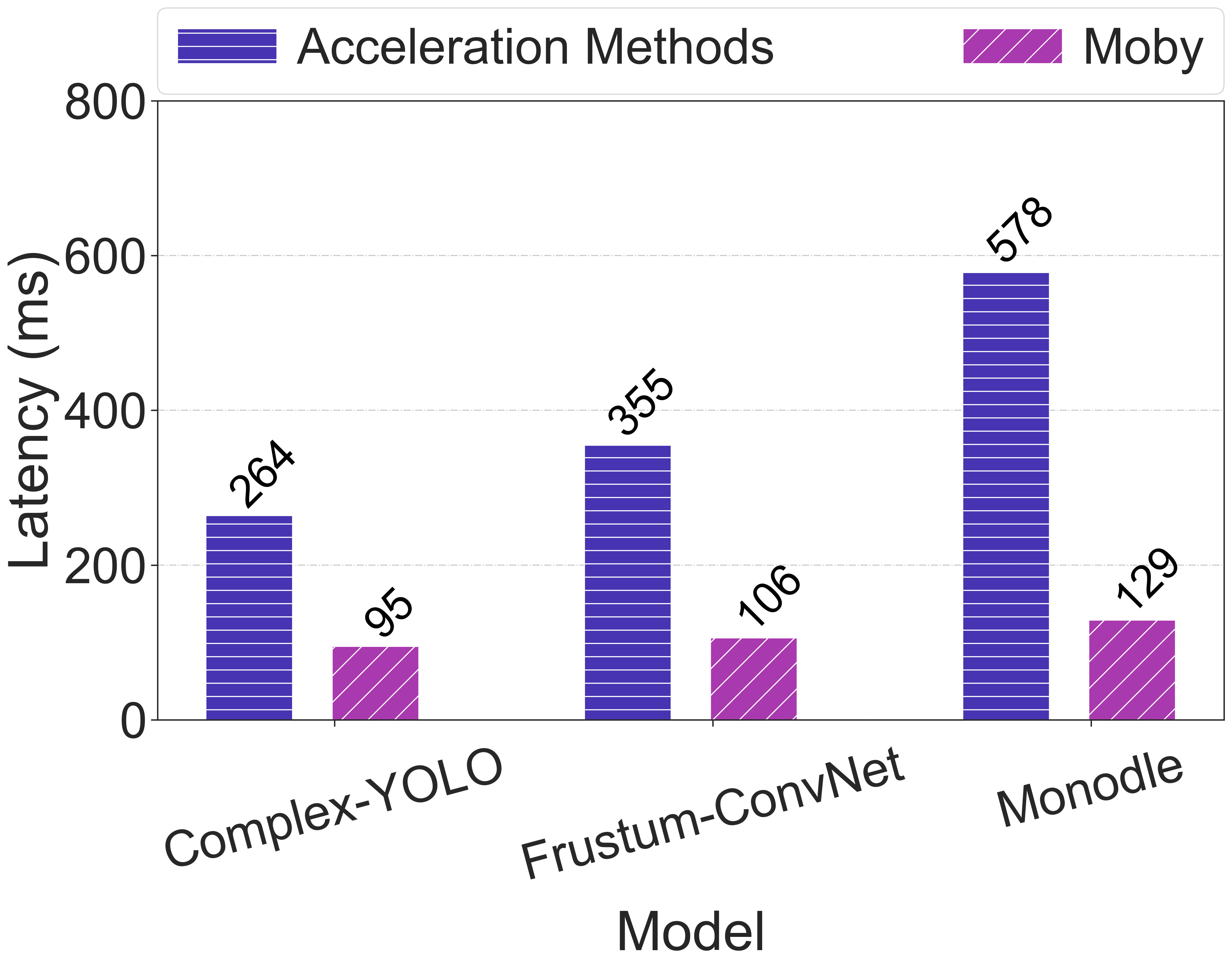}\label{fig:alternative_latency}}\subfigure[Accuracy]{\includegraphics[width=0.21\textwidth]{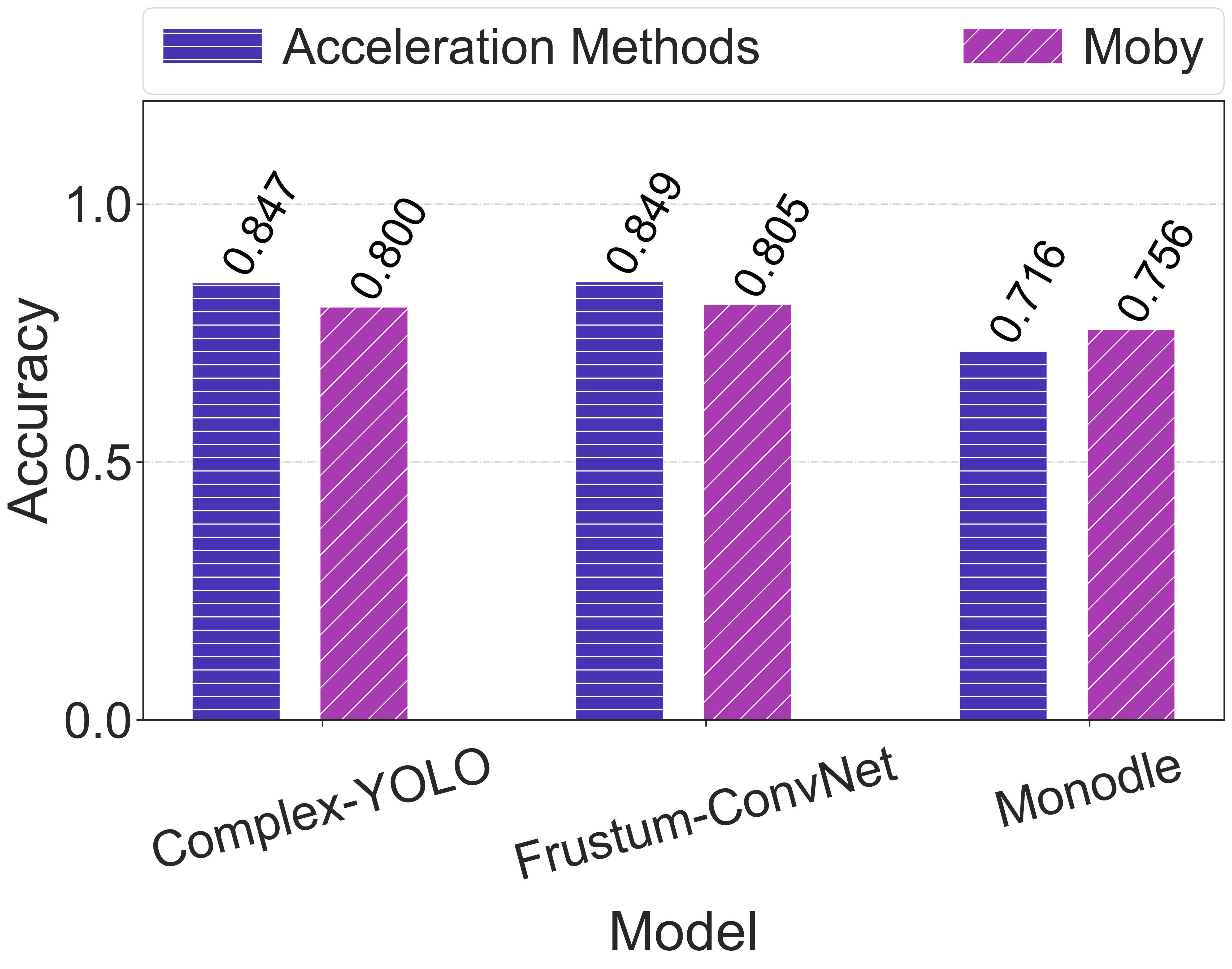}\label{fig:alternative_accuracy}}
    \vspace{-6mm}
    \caption{Comparison of \sys and alternative acceleration methods.}
    \vspace{-5mm}
\end{figure}

\subsubsection{Acceleration Methods.} 
Next, we compare \sys to three methods that accelerate 3D object detection in different ways. 

\begin{itemize}[noitemsep,topsep=0pt,parsep=0pt,partopsep=0pt,leftmargin=*]
    \item \textbf{\em Complex-YOLO} \cite{simony2018complex}: It converts point cloud data to birds-eye-view RGB maps and uses a light-weight YOLO-based architecture to predict 3D boxes. It is reported to achieve 50~FPS on an NVIDIA TitanX GPU \cite{simony2018complex}.
    \item \textbf{\em Frustum-ConvNet} \cite{wang2019frustum}: A fusion-based approach that utilizes 2D region proposals to narrow down the 3D space for {acceleration}.
    \item \textbf{\em Monodle} \cite{ma2021delving}: State-of-the-art image-based 3D detection approach that only use monocular images as input.
\end{itemize}
For a fair comparison, we also run \sys fully onboard for the anchor frames {since these methods here do not offload to cloud}, and use the same 3D detection method as the respective baseline.

Based on Fig.~\ref{fig:alternative_latency}, we observe that \sys shows significant latency improvement against the three alternative approaches. 
Specifically, \sys cuts down latency by 64.0\% compared to Complex-YOLO with only minor accuracy loss,
and the same advantage still holds against the fusion-based Frustum-ConvNet.
Notably, compared with the image-based approach Monodle, \sys reduces the latency by 77.6\% while improving accuracy by 5.5\%.
It is worth mentioning that \sys's accuracy is hindered by Monodle's performance compared to other approaches, as its inaccurate output indirectly affects the transformation accuracy of \sys. 
The relatively low accuracy of Monodle is attributed to its inability to detect locations of distant objects, which is common among image-based acceleration methods for 3D object detection \cite{qian20223d, you2019pseudo}.
We also tried another two image-based methods, Deep3DBox \cite{mousavian20173d} and Pseudo-LiDAR++ \cite{you2019pseudo}. However, they take 2834ms and 5889ms, respectively, which are too slow to execute on edge devices.

\subsection{Microbenchmarks}
\label{sec:components_analysis}
\begin{figure}[t]
\begin{minipage}{.495\linewidth}
    \LARGE
    \resizebox{1\columnwidth}{!}{
    \renewcommand{\arraystretch}{2}
        \begin{tabular}{c|ccc}
        \toprule[1.5pt]
        \textbf{Components} &\textbf{Accuracy} & \textbf{\makecell[c]{Latency\\(ms)}} & \textbf{\makecell[c]{On-board \\ Latency (ms)}}\\
        \midrule
        \rowcolor{gray!20}
        TRS & 0.762 & 88.44 & 88.44 \\
        \makecell[c]{TRS+FOS} & 0.787 & 112.06 & 89.45\\
        \makecell[c]{TRS+FOS+TBA} & 0.814 & 99.23 & 76.29 \\
        \bottomrule[1.5pt]
        \end{tabular}
    }
    \captionof{table}{Impact of each design component.}
    \label{tab:component_analysis}
\end{minipage}
\begin{minipage}{.495\linewidth}
    \includegraphics[width=1\textwidth]{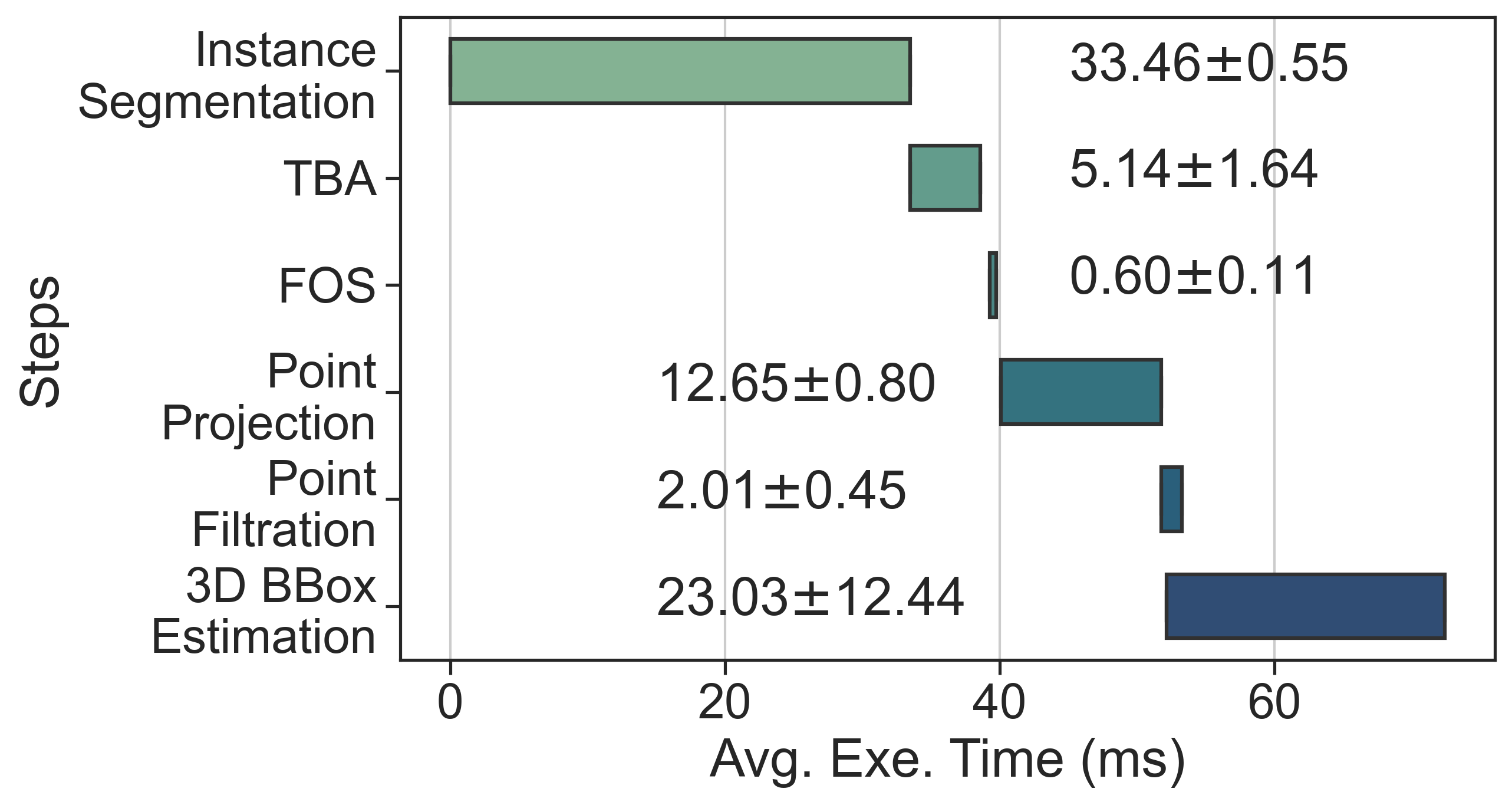}
    \vspace{-8mm}
    \caption{The avg. execution time of key steps over 300 runs.
    }
    \label{fig:latency_breakdown}
\end{minipage}
\vspace{-6mm}
\end{figure}

\noindent\textbf{Contribution of Each Component.}
We evaluate the contribution of each key design component in Moby. 
Starting with only 2D-to-3D transformation (TRS) to generate 3D results, 
we incrementally add each component, namely Frame Offloading Scheduler (FOS), and Tracking-based Association (TBA).
Table~\ref{tab:component_analysis} shows the results.
We observe that TRS can already achieve decent accuracy, proving the effectiveness of 2D-to-3D transformation.
With FOS, accuracy is improved as some frames are offloaded to server for inference. 
Offloading increases the end-to-end latency while on-board latency remains unchanged.
TBA further improves accuracy by using previous detection results to estimate more accurately.
The slightly drop of the end-to-end and on-board latency is because TBA reduces the computation overhead of TRS if boxes are successfully associated. 


\noindent\textbf{Overheads.}
We measure the execution times of key steps of \sys and report the average in Fig.~\ref{fig:latency_breakdown}.
Instance segmentation takes the longest, accounting for 43.9\% of the onboard latency, followed by 3D bounding box estimation and point projection with 30.1\% and 16.6\%, respectively, because these two steps involve multiple matrix multiplications.
TBA and FOS have negligible latency, taking only 5.14ms and 0.60ms. Notably, Point filtration has only 2.01ms, proving the efficiency of Algorithm~\ref{algo:points_filtration}.

\subsection{Sensitivity Analysis}
\label{sec:microbenchmarks}
\begin{figure}[t]
    \centering
    \subfigure[Iteration number vs. Accuracy]{\includegraphics[width=0.12\textwidth]{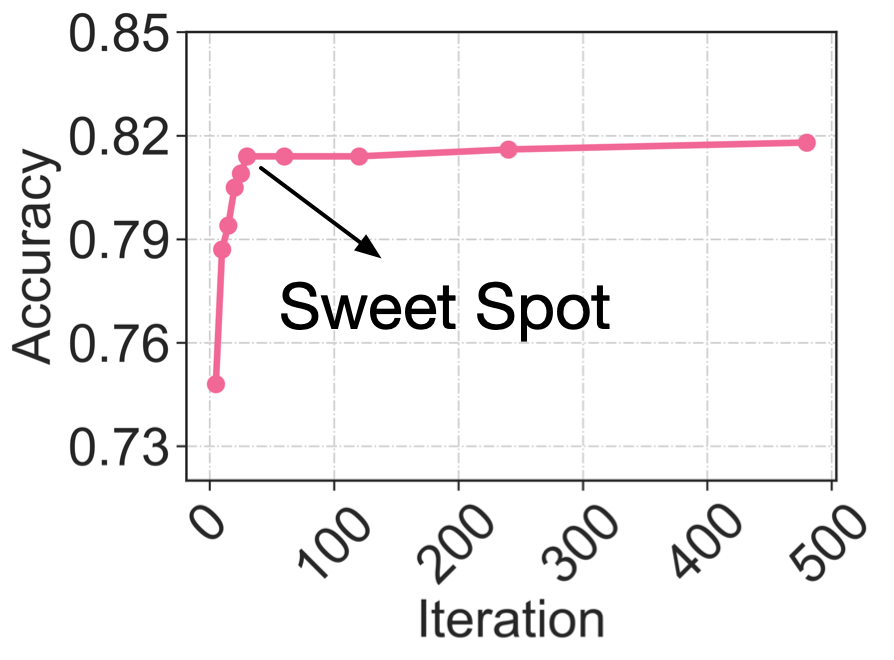}\label{fig:iter_acc}}\subfigure[Iteration number vs. Latency]{\includegraphics[width=0.12\textwidth]{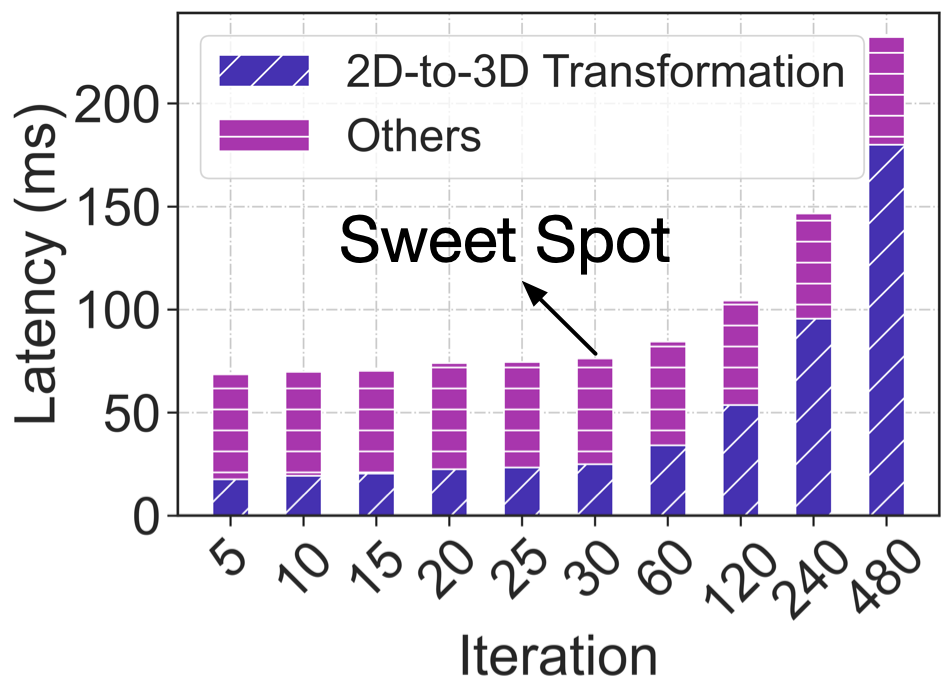}\label{fig:iter_latency}}\subfigure[Association Criterion vs. Accuracy]{\includegraphics[width=0.12\textwidth]{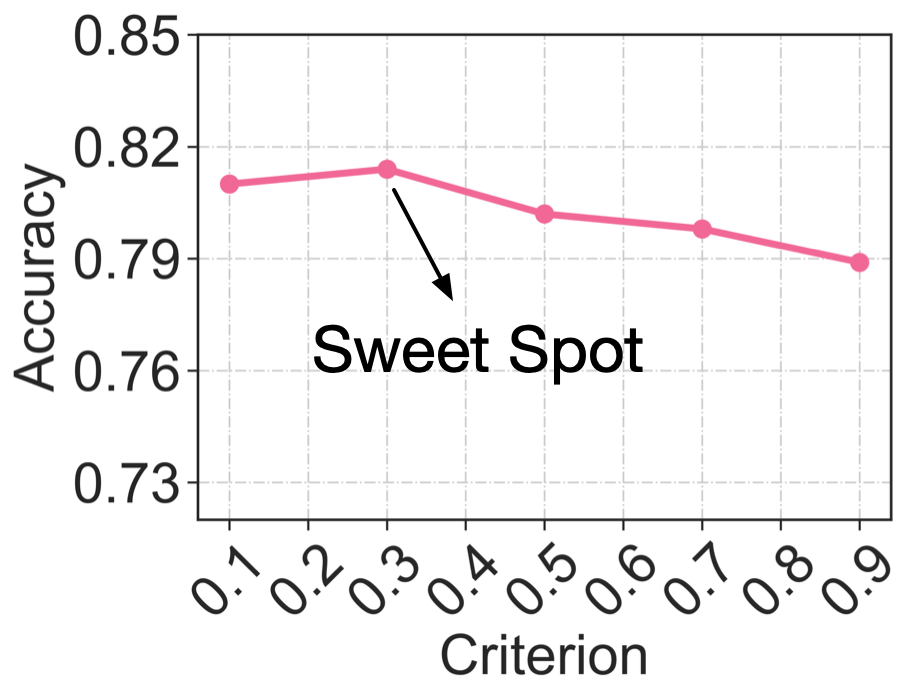}\label{fig:ass_acc}}\subfigure[Association Criterion vs. Latency]{\includegraphics[width=0.12\textwidth]{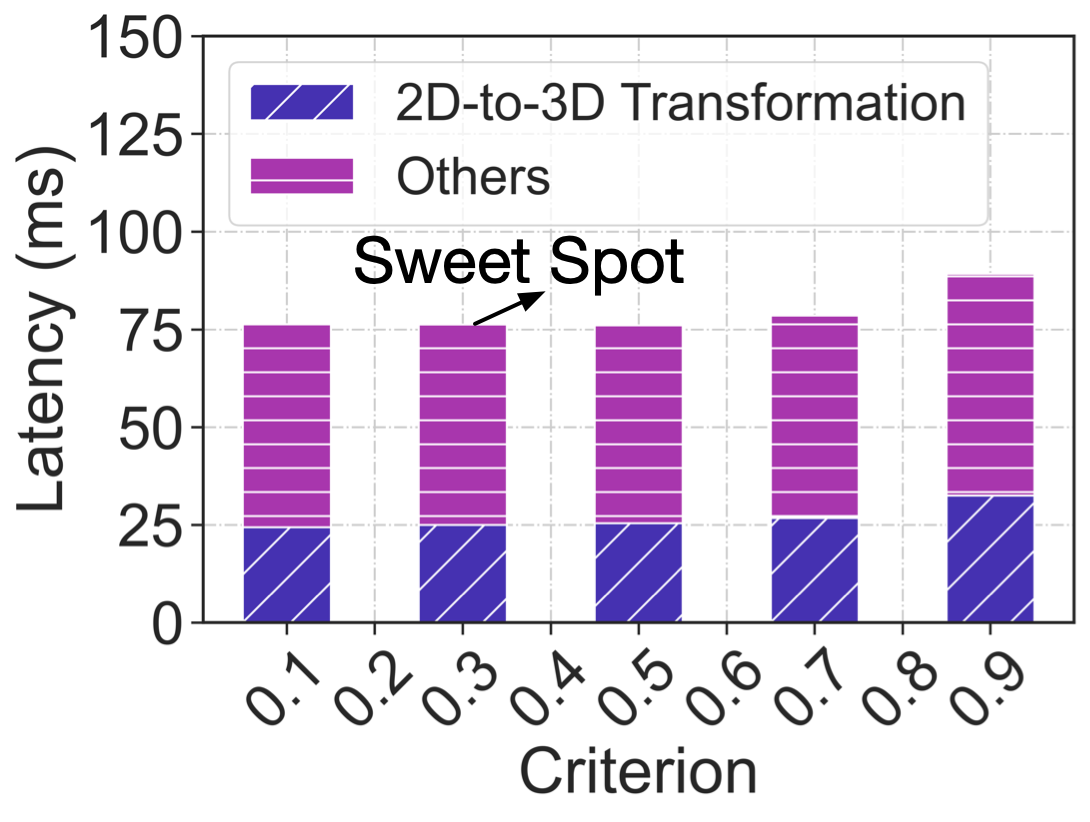}\label{fig:ass_latency}}\vspace{-5mm}\caption{The impact of key parameters. Note that here we only report the on-board latency for clearer comparison.}\label{fig:microbenchmark}\vspace{-4mm}
\end{figure}

We also benchmark \sys by changing its key parameters.

\noindent\textbf{Number of Iterations in RANSAC.}
The number of iterations in RANSAC is crucial in bounding box estimation and   affects Moby's accuracy and on-board latency, as shown in Fig.~\ref{fig:iter_acc} and Fig.~\ref{fig:iter_latency}. 
A small number reduces latency but may lead to premature or inaccurate surface recognition, while a larger number improves robustness at the expense of computation overhead. We find that 30 iterations strike an appropriate balance between accuracy and computation cost.

\noindent\textbf{Sensitivity to Association Criterion.}
Fig.~\ref{fig:ass_acc} and Fig.~\ref{fig:ass_latency} test the impact of association criterion, i.e., the IoU threshold used to associate two bounding boxes in 2D space as described in \cref{sec:kf_based_tracking}.
A larger criterion makes it harder to associate bounding boxes and cannot effectively utilize previous 3D results.
We empirically choose 0.3 as the association criterion, as the accuracy gain diminishes after it is greater than 0.3.

\subsection{System Efficiency}
Finally, we evaluate the system efficiency of \sys.

\noindent\textbf{Energy Consumption.}
We measured \sys's energy consumption using Tegrastats Utility \cite{tegrastats} for 2 minutes with a 10Hz sampling rate. Moby consumes the least power compared to baselines with 73\% savings on average. 
Its power consumption is only 24.2\% of PointPillar, and reduces GPU and CPU power by 74.3\% and 77.3\%, respectively.

\begin{figure}[t]    
        \centering
        \subfigure[Energy consumption]{\includegraphics[width=0.238\textwidth]{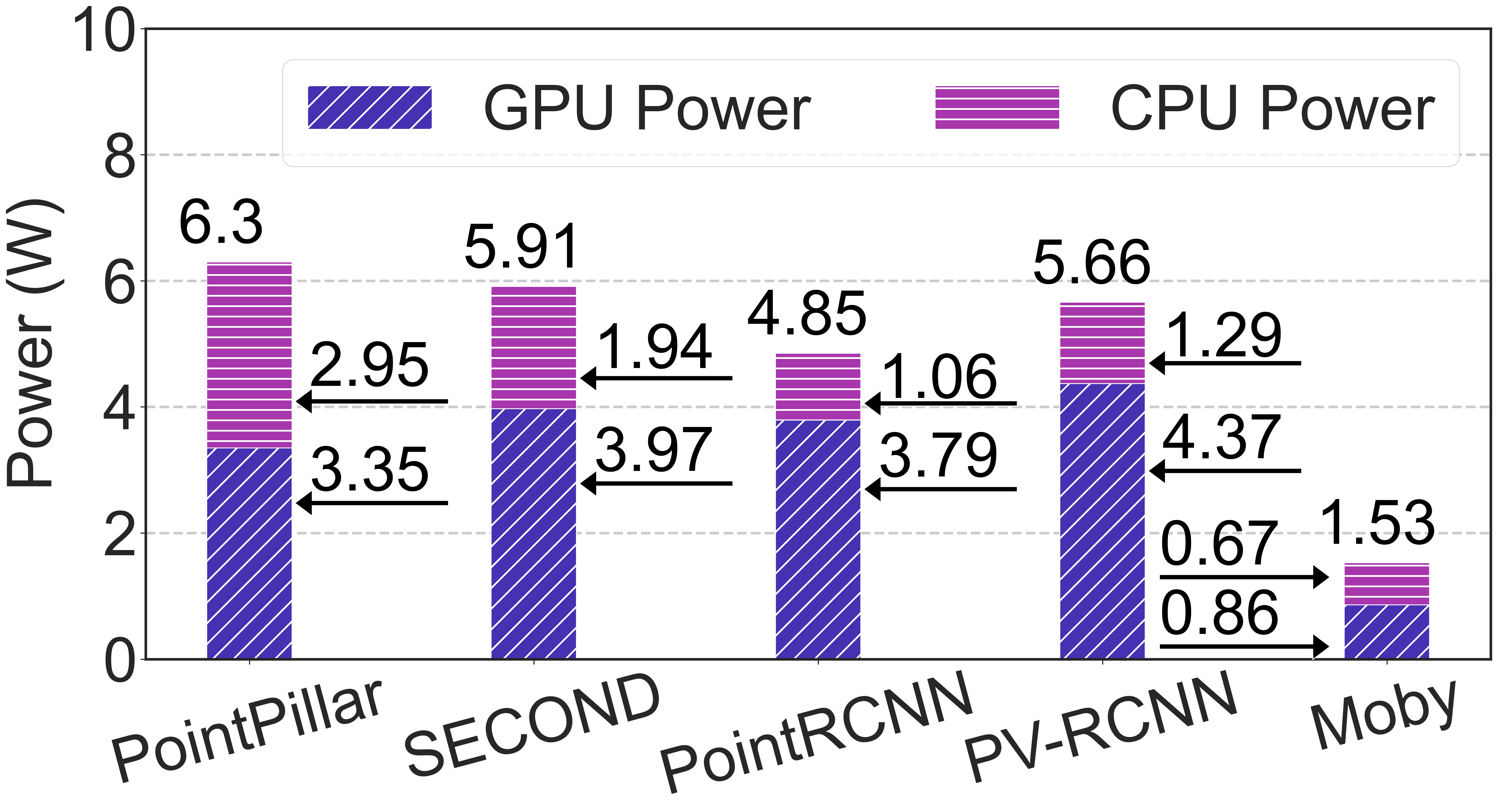}\label{fig:energy}}\subfigure[Memory footprint]{\includegraphics[width=0.238\textwidth]{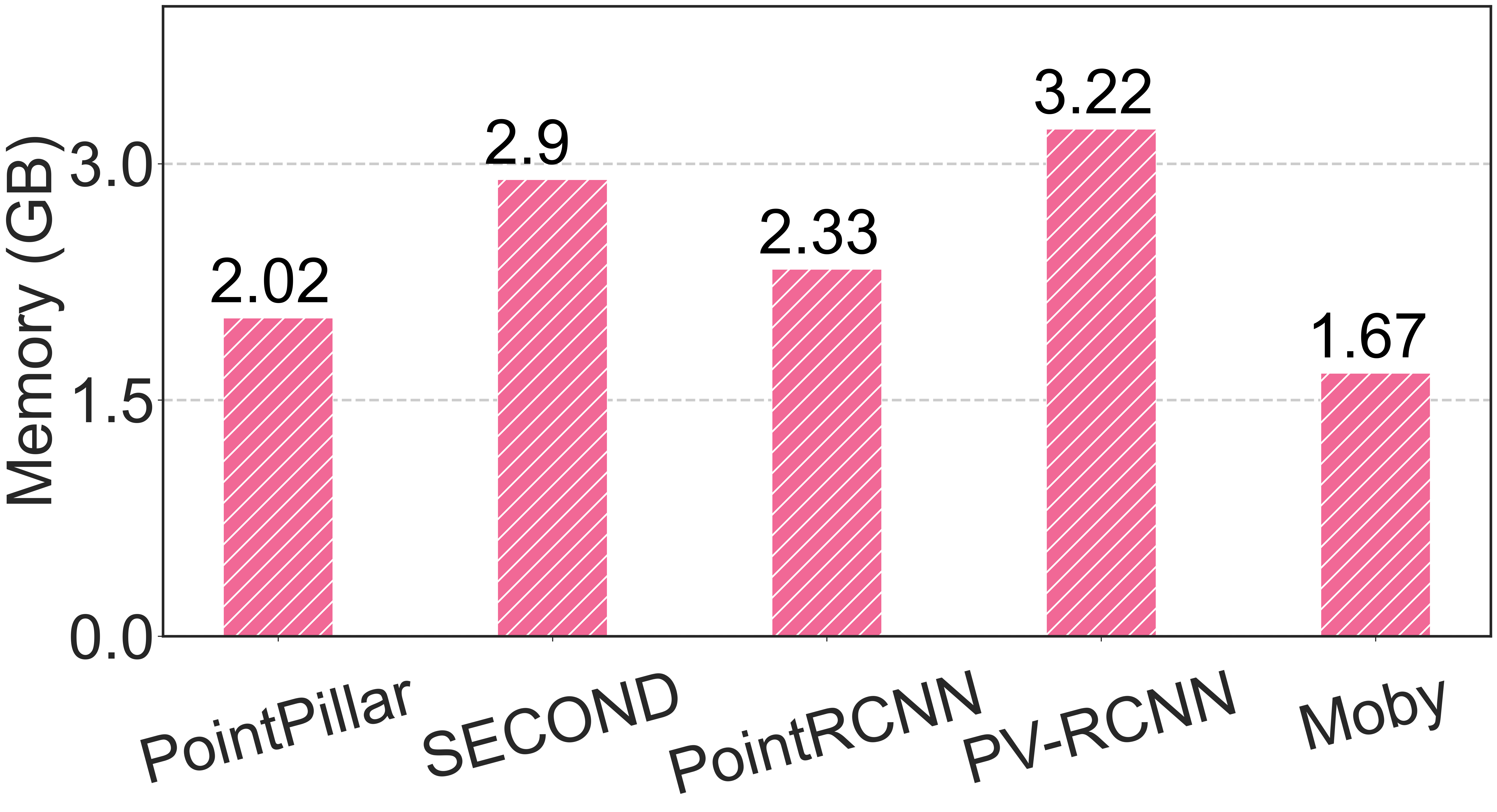}\label{fig:memory}}
        \vspace{-5mm}
        \caption{System efficiency.}
        \vspace{-3mm}
\end{figure}

\noindent\textbf{Memory Footprint.}
We measure the memory footprint of \sys and the results are shown in Fig.~\ref{fig:memory}.
\sys's overall memory reduction ranges from 17.3\% to 48.1\%,
which is expected as it only needs to load a 2D detection model into the memory that is much smaller than 3D object detection models.
\sys shows its advantage in deployment on the wimpy edge devices, such as Jetson Nano \cite{nano} that only has 4GB memory shared by both CPU and GPU.

\section{Related Work}
\label{sec:relatedwork}

\noindent\textbf{On-device Inference Acceleration.}
Lots of efforts have been devoted to accelerating DNN inference on edge devices. 
Model compression techniques reduce the model size and accelerate computation by network pruning \cite{han2015deep, he2017channel, liu2018rethinking}, quantization \cite{jacob2018quantization, courbariaux2016binarized}, and knowledge distillation \cite{hinton2015distilling, mullapudi2019online}.
Hardware-based methods design specialized accelerators for inference pipelines using FPGA \cite{umuroglu2017finn, zhang2015optimizing} or ASIC \cite{chen2014dadiannao, wang2021spatten}.
The last line of work is software-based methods \cite{tang2022torchsparse, polly, yi2020eagleeye, chen2015glimpse, jiang2021flexible}. 
\sys falls under the software-based category and is orthogonal to the existing work.

\noindent\textbf{Edge-Cloud Collaboration.}
Besides on-device inference, the emergence of edge devices has also sparked research on edge-cloud collaboration.
For example, filter-based methods \cite{zhang2021elf, li2020reducto, deng2022geryon} investigate how to prune the large data volume before transmission.
Partition-based methods \cite{zhang2020towards, kang2017neurosurgeon, jeong2018ionn} explore partitioning the DNN models across edge and cloud. 
Feedback-based methods \cite{chen2015glimpse, du2020server, nigade2021better} rely on the feedback from server to assist local processing on the edge.
Unlike existing work that all targets the 2D domain, our work explores edge-cloud collaboration for more compute-intensive 3D detection task that poses severer challenges to deploy on edge.
\section{Conclusion}
\label{sec:conclusion}

We present \sys, a novel framework for fast and accurate 3D object detection on edge devices. It leverages the 2D-to-3D transformation to significantly reduce the computational resource demand of inference, and judiciously offload the anchor frames to the cloud to maintain accuracy. 
Extensive evaluation shows that \sys reduces end-to-end latency by up to 91.9\% with little accuracy drop.

\bibliographystyle{ACM-Reference-Format}
\clearpage
\bibliography{main}
\end{document}